\documentclass[lettersize,journal]{IEEEtran}
\usepackage{amsmath,amsfonts, amssymb}
\usepackage{algorithmic}
\usepackage{algorithm}
\usepackage{array}
\usepackage[caption=false,font=normalsize,labelfont=sf,textfont=sf]{subfig}
\usepackage{textcomp}
\usepackage{stfloats}
\usepackage{url}
\usepackage{verbatim}
\usepackage{graphicx}
\usepackage{cite}
\usepackage{booktabs}
\usepackage{hyperref}
\usepackage{threeparttable}

\begin{document}

\title{CompGS++: Compressed Gaussian Splatting for Static and Dynamic Scene Representation}

\author{Xiangrui~Liu,~Xinju~Wu,~Shiqi~Wang~\IEEEmembership{Senior Member,~IEEE,}\\Zhu~Li,~\IEEEmembership{Senior Member,~IEEE,}~and~Sam~Kwong,~\IEEEmembership{Fellow,~IEEE}
\thanks{X. Liu, X. Wu and S. Wang are with the Department of Computer Science,
City University of Hong Kong, Hong Kong, China (e-mail: xiangrliu3-c@my.cityu.edu.hk; xinjuwu2-c@my.cityu.edu.hk; shiqwang@cityu.edu.hk). Z. Li is with the Department of Computer Science and Electrical Engineering, University of Missouri–Kansas City, Kansas City, MO 64110 USA (e-mail: zhu.li@ieee.org). S. Kwong is with the Department of Computing and Decision Science, Lingnan
University, Hong Kong, China (e-mail: samkwong@ln.edu.hk).}
\thanks{S. Wang is the corresponding author.}
}

\markboth{Journal of \LaTeX\ Class Files,~Vol.~14, No.~8, August~2021}%
{Shell \MakeLowercase{\textit{et al.}}: A Sample Article Using IEEEtran.cls for IEEE Journals}

\maketitle

\begin{abstract}
Gaussian splatting demonstrates proficiency for 3D scene modeling but suffers from substantial data volume due to inherent primitive redundancy.
To enable future photorealistic 3D immersive visual communication applications, significant compression is essential for transmission over the existing Internet infrastructure. 
Hence, we propose Compressed Gaussian Splatting (CompGS++), a novel framework that leverages compact Gaussian primitives to achieve accurate 3D modeling with substantial size reduction for both static and dynamic scenes. 
Our design is based on the principle of eliminating redundancy both between and within primitives. 
Specifically, we develop a comprehensive prediction paradigm to address inter-primitive redundancy through spatial and temporal primitive prediction modules. 
The spatial primitive prediction module establishes predictive relationships for scene primitives and enables most primitives to be encoded as compact residuals, substantially reducing the spatial redundancy.
We further devise a temporal primitive prediction module to handle dynamic scenes, which exploits primitive correlations across timestamps to effectively reduce temporal redundancy.
Moreover, we devise a rate-constrained optimization module that jointly minimizes reconstruction error and rate consumption.
This module effectively eliminates parameter redundancy within primitives and enhances the overall compactness of scene representations.
Comprehensive evaluations across multiple benchmark datasets demonstrate that CompGS++ significantly outperforms existing methods, achieving superior compression performance while preserving accurate scene modeling.
Our implementation will be made publicly available on GitHub to facilitate further research.
\end{abstract}

\begin{IEEEkeywords}
Gaussian splatting, spatial primitive prediction, temporal primitive prediction, rate-constrained optimization, static scene representation, dynamic scene representation, compression.
\end{IEEEkeywords}

\section{Introduction}
\IEEEPARstart{T}{he} recently proposed Gaussian splatting (3DGS)~\cite{kerbl20233d}, renowned for its proficient modeling capabilities, has catalyzed advancements in realms of 3D representation and reconstruction\cite{lei2025gaussnav, qu2024z, yin2025ms, gableman2024incorporating, chen2024s, ramirez2024deep}.
Unlike implicit representation paradigm~\cite{mildenhall2021nerf}, 3DGS~\cite{kerbl20233d} leverages explicit 3D Gaussians as primitives to represent 3D scenes and employs splatting rasterizer~\cite{zwicker2002ewa} rather than ray casting for rendering, striking an excellent trade-off between reconstruction fidelity and computational complexity. 
Subsequent advances have further broadened the practicability of 3DGS~\cite{kerbl20233d} through improvement in rendering quality~\cite{huang20242d, hamdi2024ges, wang2024tangram, li20243d, qu2024disc} and scalability to dynamic scenes~\cite{wu20244d, kratimenos2023dynmf, li2024spacetime, yang2023real,luiten2024dynamic, sun20243dgstream, gao2024hicom  }. 
However, both the original 3DGS and successive works necessitate numerous 3D Gaussians to achieve high modeling precision, resulting in millions of primitives for real-world scenes.
Such massive data volume incurs significant challenges for storage and transmission, thereby prompting urgent demands for effective compression techniques.

Prior works~\cite{navaneet2025compgs, niedermayr2024compressed, lee2024compact, girish2024eagles, liu2024compgs, chen2024hac} to compressing Gaussian splatting representations follow two strategies: reducing the quantity of 3D Gaussians and minimizing the data footprint of each 3D Gaussian.  
Specifically, several methods~\cite{navaneet2025compgs, niedermayr2024compressed, lee2024compact, girish2024eagles} employ heuristic pruning to cull redundant 3D Gaussians that contribute insignificant to scene modeling. 
Quantization is then used to substitute parameters of the remaining primitives with discrete codewords for further size reduction.
However, such rudimentary pruning strategies risk over-aggressive primitive removal and potentially compromise scene modeling accuracy.
Furthermore, quantization on individual 3D Gaussians ignores inter-primitive redundancy.
More sophisticated works~\cite{liu2024compgs, chen2024hac} integrate prediction mechanisms into the compression pipeline, capitalizing on the inter-primitive correlations to eliminate redundancy.
Yet, the spatial correlations among 3D Gaussians remain insufficiently explored, suggesting considerable potential for improvement. 
Moreover, existing compression methods only focus on static scenes, leaving the compression challenge in dynamic scenes unresolved.

\begin{figure}[t]
  \centering
  \includegraphics[width=1.\linewidth]{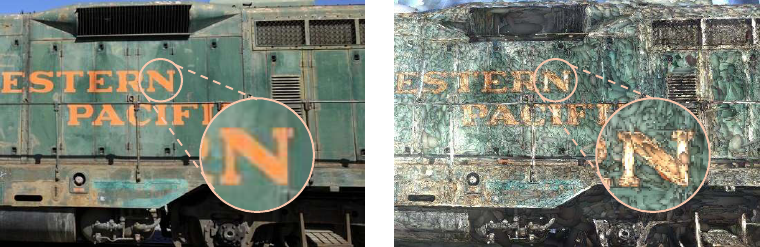}
  \caption{Illustration of spatial correlations of 3D Gaussians. \textbf{Left}: Scene visualization having enlarged view of objects with inherent continuity. \textbf{Right}: Demonstration of primitive correlations arising from the scene continuity.}
  \label{fig::spatial_correlation_visualization}
\end{figure}

\begin{figure}[t]
  \centering
  \includegraphics[width=0.85\linewidth]{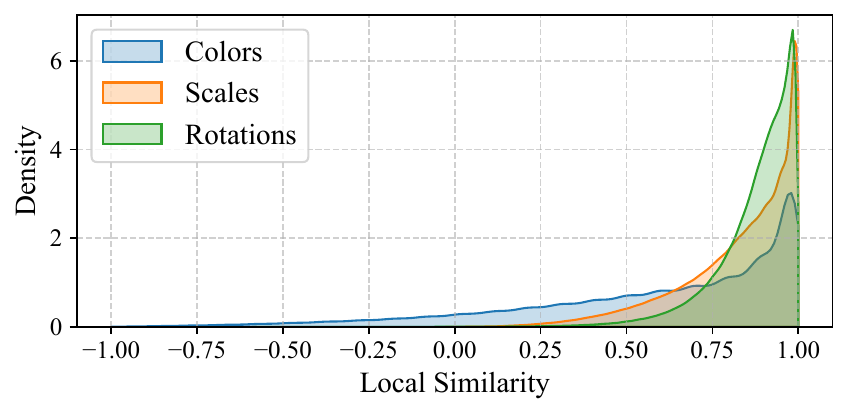}
  \caption{Statistical analysis of spatial correlations between 3D Gaussians, wherein spatial correlations are measured by the average cosine distances between a 3D Gaussian and its 20 nearest neighbors.}
  \label{fig::spitial_correlation_statistic}
\end{figure}

This paper presents Compressed Gaussian Splatting (CompGS++), a holistic compression framework for compact 3D representation of both static and dynamic scenes.  
Specifically, to foster compact Gaussian splatting representation for static scenes, we propose a spatial primitive prediction module to reduce redundancy among primitives.
The insight behind our prediction module is rooted in the spatial correlations of 3D Gaussians.
As shown in Fig.~\ref{fig::spatial_correlation_visualization} and Fig.~\ref{fig::spitial_correlation_statistic}, the inherent continuity of scenes contributes to pronounced correlations in 3D Gaussians, allowing the establishment of predictive relationships among these primitives. 
To achieve the prediction, we craft a hybrid primitive structure that enlists a limited amount of anchor primitives as references to predict the remaining coupled primitives.
The sparse distribution of anchor primitives allows for the incorporation of ample reference information while avoiding significant coding overhead.
Meanwhile, coupled primitives can be effectively predicted by anchor primitives owing to the potent inter-primitive correlations, and thus, necessitate only compact residues to compensate for prediction errors.
Based on the hybrid primitive structure, the proposed prediction module can effectively eliminate inter-primitive redundancy and compactly represent 3D Gaussians.
To further optimize the parameter efficiency of these primitives, we propose a rate-distortion optimization module that cultivates rate regularization reflecting the coding costs of primitives. 
This optimization module jointly minimizes reconstruction distortion and primitive coding costs, striking an optimal rate-distortion trade-off in 3D modeling.

\begin{figure}[t]
  \centering
  \includegraphics[width=0.9\linewidth]{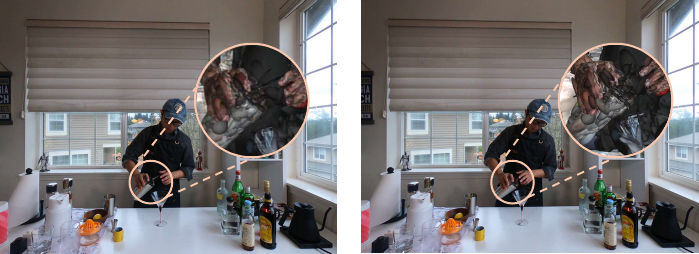}
  \caption{Visualization of temporal correlations between 3D Gaussians. \textbf{Left}: Previous P-scene with enlarged view of 3D Gaussians in the dynamic area. \textbf{Right}: Subsequent P-scene with visualized 3D Gaussians in the same dynamic area, demonstrating notable temporal correlations of 3D Gaussians.}
  \label{fig::temporal_correlation_visualization}
\end{figure}

\begin{figure}[t]
  \centering
  \includegraphics[width=0.95\linewidth]{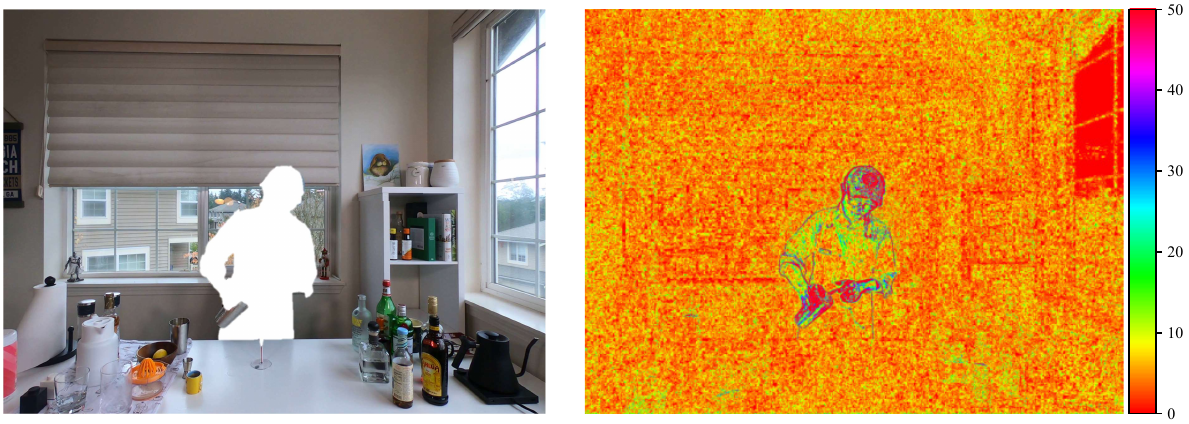}
  \caption{Illustration of temporal shifts in ambient illumination. \textbf{Left}: Static regions of the current P-scene. \textbf{Right}: Difference between the current and subsequent P-scenes, revealing subtle temporal ambient illumination variations in static regions.}
  \label{fig::temporal_correlation_statistic}
\end{figure}

We further elevate our framework beyond static scene compression to achieve compact representations of dynamic scenes.
We represent sequential scenes by the streamable configuration~\cite{sun20243dgstream, gao2024hicom}, with a combination of intra scenes (I-scenes) and predicted scenes (P-scenes). 
For I-scenes, we employ our established static scene compression methodology as the foundational processing paradigm.
The P-scenes are sequentially processed via our specialized dynamic compression paradigm that capitalizes on the temporal correlations among 3D Gaussians.
Specifically, as illustrated in Fig.~\ref{fig::temporal_correlation_visualization}, successive P-scenes showcase consistent motion patterns characterized by continuity and locality, revealing profound temporal correlations among 3D Gaussians.
Motivated by such correlations, we develop a temporal primitive prediction module that exploits reference primitives from preceding scenes to craft the primitive constitution of the current scene.
Only succinct temporal residues are cultivated for the P-scenes to perform temporal deformation on reference primitives, thereby facilitating compact scene representations. 
Moreover, inspired by the locality of scene dynamics, we disentangle reference primitives into dynamic and static categories in the prediction module.
Dynamic primitives undergo comprehensive adjustment on both geometry and appearance attributes.
Conversely, the static ones are subtly altered only in appearance attributes to accommodate the temporal shifts of ambient lighting shown in Fig.~\ref{fig::temporal_correlation_statistic}.
The proposed disentangling strategy can judiciously refine motion modeling by applying geometry warping on a selective set of primitives, unlike extant works~\cite{sun20243dgstream, gao2024hicom} that indiscriminately apply such warping over all primitives.
Additionally, we devise a temporal adaptive control mechanism to improve the dynamic-static primitive disentanglement.
This control mechanism facilitates adaptive state transitions between dynamic and static primitives based on meticulously designed indicators, enabling precise representation of both moving and stationary scene elements.

The main contributions of our work are summarized as follows:
\begin{itemize}
\item We propose a holistic compression framework, namely CompGS++, to furnish compact 3D representations for both static and dynamic scenes. 
By exploiting redundancy-eliminated primitives to characterize 3D scenes, CompGS++ achieves superior compression performance across leading static and dynamic benchmarks. 
\item We develop a novel spatial primitive prediction module for static scenes that effectively minimizes inter-primitive redundancy.
Additionally, we propose a rate-constrained optimization module that utilizes rate costs as regularization to enhance the compactness of primitive parameters.
\item We design a compact representation paradigm for dynamic scenes, incorporating a temporal primitive prediction module to efficiently eliminate primitive redundancy between consecutive scenes. 
Furthermore, we propose a temporal adaptive control mechanism to improve the efficacy of dynamic modeling, thereby yielding accurate and compact scene representations. 
\end{itemize}

\section{Related Work}
\subsection{Static Scene Representation by Gaussian Splatting}
The groundbreaking 3DGS~\cite{kerbl20233d} is distinguished from previous implicit neural radiance fields~\cite{mildenhall2021nerf} by the utilization of explicit primitives and accelerated splatting rasterization.
Specifically, 3DGS~\cite{kerbl20233d} leverage 3D Gaussians as primitives, with locations and shapes determined by geometry attributes, i.e., means and covariances.
Furthermore, appearance attributes, including opacities and colors, are integrated into 3D Gaussians to portray scene colors. 
Such explicit primitive design allows 3DGS~\cite{kerbl20233d} to bypass the laborious ray sampling pipeline~\cite{mildenhall2021nerf} and opt for a highly parallelized volume splatting rasterizer~\cite{zwicker2002ewa}.
Moreover, 3DGS~\cite{kerbl20233d} involves an adaptive control mechanism to heuristically regulate the density of 3D Gaussians, alternating between densification and pruning to create or remove primitives.

Building upon the pioneering 3DGS~\cite{kerbl20233d}, several works have obtained advancements in primitive design~\cite{huang20242d,hamdi2024ges, wang2024tangram}, optimization strategy~\cite{ye2024absgs, bulo2024revising, zhang2024pixel}, and rendering pipeline~\cite{wang2024adr, huang2024gs++}. 
Specifically, Huang et~al.~\cite{huang20242d} proposed to refine the primitive design by converting 3D Gaussians into 2D Gaussian disks, improving rendered view consistency. 
Hamdi et~al.~\cite{hamdi2024ges} devised generalized 3D Gaussians, enhancing boundary modeling accuracy by adaptively adjusting the exponent of Gaussian distribution.
Subsequent work proposed by Wang et~al.~\cite{wang2024tangram} introduced a mixture Gaussian primitive structure to meticulously capture intricate scene details. 
To improve the optimization strategies of 3DGS~\cite{kerbl20233d}, Ye et~al.~\cite{ye2024absgs} integrated a homo-directional gradient correction for adaptive density control, which can effectively mitigate gradient collisions and improve the efficacy of densified 3D Gaussians. 
Bul\`{o} et~al.~\cite{bulo2024revising} suggested involving pixel-level rendering distortions in adaptive control to create essential primitives. 
Additionally, Zhang et~al.~\cite{zhang2024pixel} devised an improved densification scheme, leveraging size regularization to address the imperfections brought by the original size-agonistic densification process.
Turning to rendering advancements, Wang et~al.~\cite{wang2024adr} proposed thread-level culling and pixel-level load balancing strategies to reduce the complexity of rasterization. 
Furthermore, Huang et~al.~\cite{huang2024gs++} analyzed projection errors arising from differentiable approximations in rendering and developed a corrected projection strategy to reduce the errors.

Despite these advancements, the foundational 3DGS~\cite{kerbl20233d} and its subsequent works necessitate numerous 3D Gaussians to uphold reconstruction quality, culminating in significant model size.

\subsection{Dynamic Scene Representation by Gaussian Splatting}
The evolution of 3DGS~\cite{kerbl20233d} has propelled its applications in dynamic scene representation (4DGS), wherein temporal variables are integrated with 3D Gaussians to encapsulate scene dynamics.
Existing 4DGS works fall into two categories: non-streamable~\cite{wu20244d, kratimenos2023dynmf, li2024spacetime, yang2023real, luiten2024dynamic} and streamable\cite{sun20243dgstream, gao2024hicom}. 
Non-streamable methods endeavor to capture the scene into a single representation model, including 3D Gaussians cultivated in a canonical space and time-varying deformation fields.
These methods embed scene dynamics into the deformation fields and transform primitives based on the deformation fields to reconstruct scenes at specific moments.
Specifically, Wu et~al.~\cite{wu20244d} designed spatiotemporal voxel planes to effectively encode temporal deformation fields for 3D Gaussians. Kratimenos et~al.~\cite{kratimenos2023dynmf} proposed to leverage parametric trajectories to guide the temporal deformation of 3D Gaussians.
To facilitate the accuracy of dynamic modeling, Li et~al.~\cite{li2024spacetime} devised to implicitly represent deformation fields by elevating the static parameters of 3D Gaussians into temporal functions.
Furthermore, Yang et~al.~\cite{yang2023real} enhanced 3D Gaussians with temporal factors to enable the evolution of 3D Gaussians over time, thereby achieving modeling of dynamic scenes.

Follow-up research~\cite{sun20243dgstream, gao2024hicom} reveals that non-streamable 4DGS methods encounter practical limitations in scalable modeling and progressive transmission.
These limitations further spur the development of streamable paradigm.
This novel paradigm chronologically tackles dynamic scenes, and thus facilitates not only incremental construction but also progressive transmission.
In this vein, Sun et~al.~\cite{sun20243dgstream} proposed a streamable framework named 3DGStream, involving a neural transformation cache module to cultivate deformation fields between two consecutive scenes. 
3D Gaussians from the preceding scene are warped based on the deformation fields to construct the subsequent scene.
Additionally, Gao et~al.~\cite{gao2024hicom} developed a Hierarchical Coherent Motion (HiCoM) framework, which segments input scenes into distinct regions by motion intensity and employs hierarchical deformation fields to intricately capture complex dynamics.

Acknowledging that motion parameters in current 4DGS methods lead to an increased volume of scene representations, it becomes imperative to devise a compact Gaussian representation scheme tailored for dynamic scene modeling.

\subsection{Compression on Gaussian Splatting Representations}
Recent works proposed to reduce the size of Gaussian splatting representations through two primary tactics: pruning to decrease the quantity of 3D Gaussians and quantization to reduce the footprint of each primitive.
Specifically, Navaneet et~al.~\cite{navaneet2025compgs} proposed to apply vector quantization on 3D Gaussians, discretizing continuous parameters into online-optimized codewords to reduce the storage footprint.
Niedermayr et~al.~\cite{niedermayr2024compressed} devised a sensitivity-based quantization framework for 3D Gaussians, wherein sensitivities measured by gradients are used to optimize the quantization codebook.
Fan et~al.~\cite{fan2024lightgaussian} proposed an importance-based pruning strategy, ranking 3D Gaussians by their rendering contributions during pruning. Meanwhile, a distillation technique is employed on spherical harmonics to alleviate the overheads of color modeling.

Subsequent works~\cite{lee2024compact, girish2024eagles, chen2024hac, liu2024compgs} have introduced neural networks into compression pipeline to further improve the compactness of Gaussian splatting representations.
Specifically, Lee et~al.~\cite{lee2024compact} observed that, within the 3DGS~\cite{kerbl20233d}, the spherical harmonics for color modeling constitute a substantial portion of the parameters. 
To address this, they devised a grid-based neural field to compactly model color attributes.
Girish et~al.~\cite{girish2024eagles} proposed to embed parameters of 3D Gaussians into compact features and recover these parameters by neural networks before rendering.
Chen et~al.~\cite{chen2024hac} proposed a context model based on binary feature grid, exploiting spatial correlations among primitives to facilitate the compression.
Liu et~al.~\cite{liu2024compgs} devised a predictive paradigm to represent most 3D Gaussians via a compact residual form. Meanwhile, a rate-constraint optimized framework is devised to improve the compactness of parameters within primitives.

Nonetheless, current methods insufficiently address redundancies in 3D Gaussian representations. 
To address this limitation, we devise an advanced spatial prediction mechanism in CompGS++.
This mechanism builds upon our preliminary work~\cite{liu2024compgs} and improves the prediction efficacy through a sophisticated contextual modeling module.
Moreover, we strengthen the rate-constrained optimization~\cite{liu2024compgs} with an enhanced entropy model, incorporating effective spital priors to improve the accuracy of rate modeling.
These improvements contribute to the superiority of CompGS++ against existing works~\cite{navaneet2025compgs, niedermayr2024compressed, lee2024compact, girish2024eagles, liu2024compgs, chen2024hac}.
Notably, CompGS++ further pioneers the compact modeling of dynamic scenes, an area unexplored by existing methods.
Through our adeptly designed temporal primitive prediction and adaptive control mechanisms, CompGS++ delivers remarkable redundancy elimination for dynamic scene representations.

\begin{figure*}[t]
    \centering
    \includegraphics[width=0.9\linewidth]{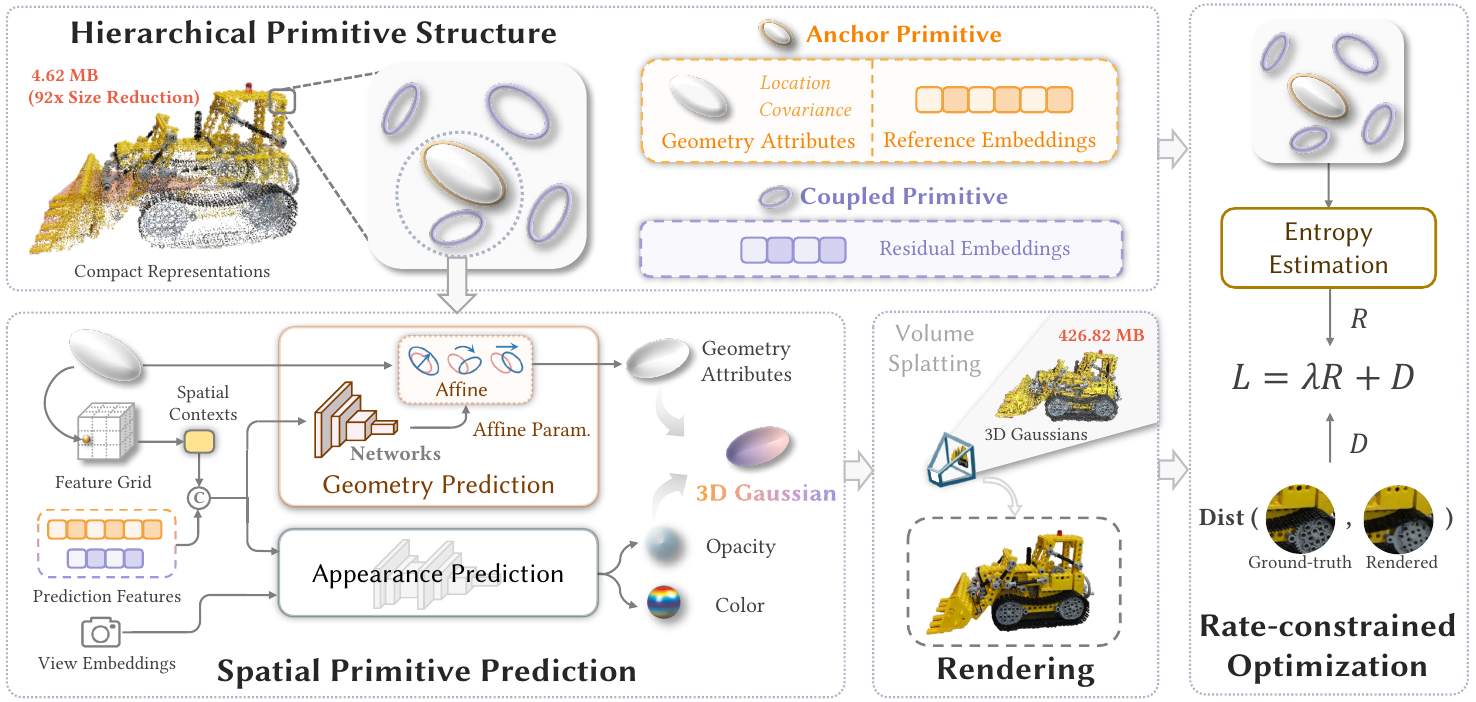}
    \caption{Overview of the proposed compact representation framework for static scenes.}
    \label{fig::static_overview}
\end{figure*}

\section{Compressed Gaussian Splatting for Static Scenes}\label{sec::static}
\subsection{Overview}
3DGS~\cite{kerbl20233d} constructs static scenes via a set of 3D Gaussians.
Hence, the essence of compressing Gaussian splatting representations lies in the compression of these primitives.
As shown in Fig.~\ref{fig::static_overview}, we propose two synergistic modules to achieve efficient compression: spatial primitive prediction to reduce the volume of primitives, and rate-constrained optimization to improve the compactness of primitive parameters. 
Specifically, we devise a spatial prediction module, capitalizing on the spatial correlations among 3D Gaussians to reduce primitive volumes.
This module is built upon a hierarchical primitive structure, encompassing anchor primitives $\Omega$ and associated coupled primitives $\Gamma$.
Each anchor primitive $\omega \in \Omega$ encapsulates geometry characteristics including location $\mu_{\omega}$ and covariance $\Sigma_{\omega}$.
Additionally, reference embeddings $f_{\omega}$ are incorporated into $\omega$ to provide reference information for prediction.
The anchor primitive $\omega$ is further associated with $K$ coupled primitives $\Gamma_{\omega} \subseteq \Gamma$, where each primitive $\gamma_k \in \Gamma_{\omega}$ is crafted with only succinct residual embeddings $f_{\gamma_k}$.
Subsequently, 3D Gaussians are efficiently predicted based on $\{\Omega, \Gamma\}$.
The prediction process entails geometry prediction to yield geometry attributes of 3D Gaussians by affine transform on $\omega$, with affine parameters predicted based on $\{f_{\omega}, f_{\gamma_k}\}$.
Concurrently, appearance attributes of 3D Gaussians are derived through appearance prediction that also hinges on $\{f_{\omega}, f_{\gamma_k}\}$.
Owing to the prediction mechanism, 3D Gaussians can be efficiently represented by a limited quantity of anchor primitives alongside numerous coupled primitives, wherein these coupled primitives can be succinctly presented by residual embeddings. 
Hence, this prediction mechanism adeptly mitigates primitive redundancy and provides compact Gaussian splatting representations.
The detailed prediction process is elaborated in Section~\ref{sec::spatial_pred}. 

Furthermore, we devise a rate-constrained optimization module to enhance parameter compactness for both anchor and coupled primitives. 
This module is crafted with a dual objective: maximizing the rendering quality of primitives while simultaneously minimizing the coding costs. 
Rendering quality is measured by the discrepancy between a training image $x$ and its counterpart $\tilde{x}$ rendered by predicted 3D Gaussians.
Meanwhile, the coding costs are determined by the parameter bitrate of the anchor and coupled primitives, assessed by a dedicated entropy model.
Herein, the optimization objective can be articulated as:
\begin{equation}\label{eq::rd_loss}
    \Omega^\star, \Gamma^\star=\mathop{\arg\min}_{\boldsymbol{\Omega}, \boldsymbol{\Gamma}}\big\{D\left(x, \tilde{x}\right)+ \lambda R\left(\Omega, \Gamma\right)\big\},
\end{equation}
where $D$ denotes the quality item and $R$ denotes the rate item. Details of the proposed rate-constrained optimization module are demonstrated in Section~\ref{sec::rate_constrained_optimization}

\begin{figure}[t]
  \centering
  \includegraphics[width=0.4\textwidth]{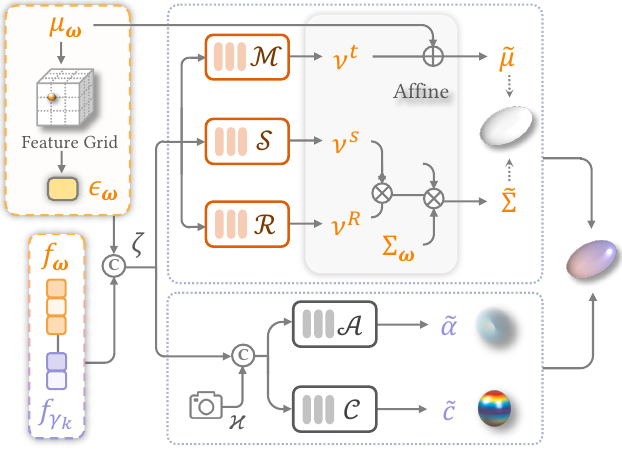}
  \caption{Illustration of the proposed spatial primitive prediction module.}
  \label{fig::spatial_pred}
\end{figure}

\subsection{Spatial Primitive Prediction}\label{sec::spatial_pred}
The spatial primitive prediction is proposed to compactly represent 3D Gaussians based on the hybrid primitive structure.
As depicted in Fig.~\ref{fig::spatial_pred}, the output 3D Gaussian $g_k$ is derived through a synergetic utilization of the coupled primitive $\gamma_k$ and its corresponding anchor primitive $\omega$.
This process involves predicting both geometry attributes (location $\tilde{\mu}$ and covariance $\tilde{\Sigma}$) and appearance attributes (color $\tilde{c}$ and opacity $\tilde{\alpha}$) for $g_k$.
Specifically, we harness residual embeddings $f_{\gamma_k}$ from $\gamma_k$ alongside $f_{\omega}$, the reference embeddings within $\omega$, to provide primitive features for the prediction.
Herein, $f_{\omega}$ offers ample reference information shared across linked coupled primitives, while $f_{\gamma_k}$ contributes supplementary details pertinent to $g_k$ for refinement.
To augment prediction priors, we further explore correlations among anchor primitives to harvest spatial contexts $\epsilon_{\omega}$. 
In this process, the location $\mu_{\omega}$ of $\omega$ is used to query $\epsilon_{\omega}$ from a learnable feature grid~\cite{chen2024hac}.
The inherent hash-encoding property of this grid enables $\epsilon_{\omega}$ to effectively capture extensive correlations among adjacent anchor primitives.
Prediction features $\zeta$ are then generated by fusing primitive features and spatial contexts, i.e., 
\begin{equation}
    \zeta = f_\omega \oplus f_{\gamma_k} \oplus \epsilon_\omega,
\end{equation}
where $\oplus$ denotes the channel-wise concatenation.
Thereafter, we propose to predict geometry attributes $\{\tilde{\mu}, \tilde{\Sigma}\}$ through an affine transform applied on $\omega$, with affine offsets $\nu$ derived from our prediction features $\zeta$, i.e.,
\begin{equation}\label{eq::affine}
    \tilde{\mu}, \tilde{\Sigma} = \mathcal{T}(\mu_{\omega}, \Sigma_{\omega} | \nu),
\end{equation}
where $\mathcal{T}$ denotes the affine transform, and $\{\mu_{\omega}, \Sigma_{\omega}\}$ denote location and covariance of $\omega$, respectively.   
The affine transform encompasses translation, scaling and rotation, allowing Eq.~(\ref{eq::affine}) to be delineated as:
\begin{equation}
    \tilde{\mu} = \mu_{\omega} + \nu^t, \quad
    \tilde{\Sigma} = Diag(\nu^s)\times \nu^R \times \Sigma_{\omega},
\end{equation}
where $\times$ denotes matrix multiplication, $\{\nu^t, \nu^s, \nu^R\}$ correspond to the translation vector, scaling vector, and rotation matrix within $\nu$, respectively, and $Diag(\nu^s)$ denotes the diagonal matrix constructed by $\nu^s$.
Accordingly, we use three lightweight linear layers to yield affine offsets from the prediction features $\zeta$, i.e.,
\begin{equation}
        \nu^t = \mathcal{M}(\zeta), \quad \nu^s = \mathcal{S}(\zeta), \quad \nu^R = \mathcal{R}(\zeta),
\end{equation}
where $\{\mathcal{M},\mathcal{S},\mathcal{R}\}$ denote the linear layers.
Moreover, we proceed to generate appearance attributes of $g_k$ in a view-dependent manner.
Specifically, view embeddings $\varkappa$, derived from the camera pose, are fused with the prediction features $\zeta$ and then passed through linear layers to predict the color $\tilde{c}$ and opacity $\tilde{\alpha}$.
This process can be formulated as:
\begin{equation}
   \tilde{c} = \mathcal{C}(\varkappa \oplus \zeta), \quad \tilde{\alpha} = \mathcal{A}(\varkappa \oplus \zeta), 
\end{equation}
where $\{\mathcal{C}, \mathcal{A}\}$ denote the linear layers for color and opacity prediction, respectively.

The proposed spatial primitive prediction commences from the compact hybrid primitives $\{\omega, \gamma_k\}$, and cultivates the 3D Gaussian $g_k$ by predicted attributes $\{\tilde{\mu}, \tilde{\Sigma}, \tilde{c}, \tilde{\alpha}\}$, thereby achieving compact representation of 3D scenes.

\begin{figure}[t]
  \centering
  \includegraphics[width=0.45\textwidth]{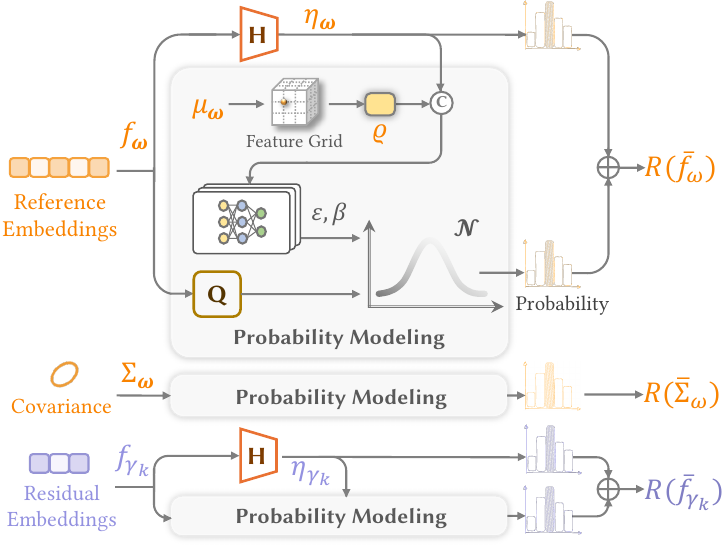}
  \caption{Illustration of the entropy model within our rate-constrained optimization module.}
  \label{fig::entropy_model}
\end{figure}

\begin{figure*}[t]
  \centering
  \includegraphics[width=0.9\linewidth]{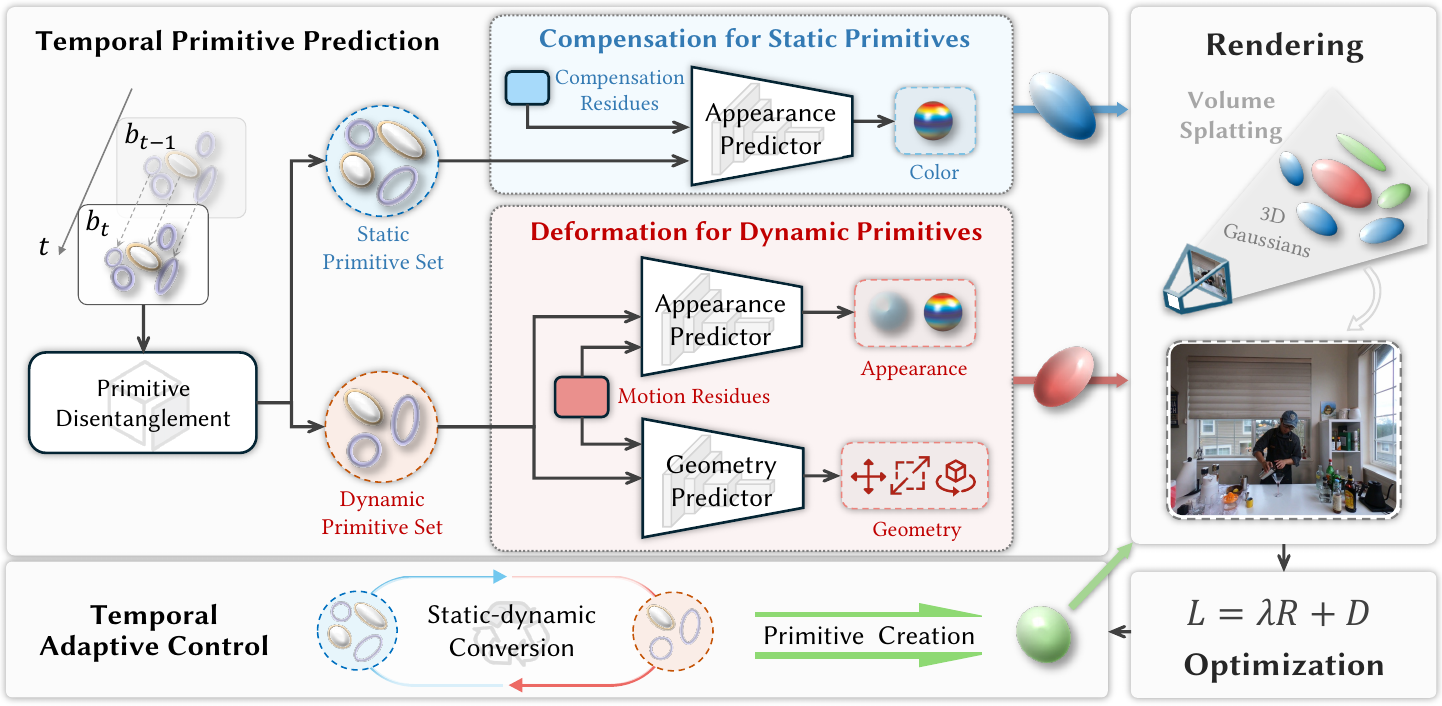}
  \caption{Overview of the proposed compact representation framework for dynamic scenes.}
  \label{fig::dynamic_overview}
\end{figure*}

\subsection{Rate-constrained Optimization}\label{sec::rate_constrained_optimization}
The rate-constrained optimization module is proposed to improve the compactness of parameters within the anchor and coupled primitives by jointly minimizing rendering distortion and coding costs.
Within the optimization objective illustrated in Eq.~(\ref{eq::rd_loss}), the rate item $R(\Omega, \Gamma)$ fundamentally hinges on the primitive parameter entropy $H(\Omega, \Gamma)$. 
However, $H(\Omega, \Gamma)$ is intractable to be calculated since the ground-truth parameter distributions are unknown, thus necessitating an estimation of the distributions.
Accordingly, the rate item in the optimization embodies both the parameter entropy and estimation bias, i.e.,
\begin{equation}
\begin{aligned}
    R(\Omega, \Gamma) &= H(\Omega, \Gamma) + \mathbb{KL}\left[p_g(\Omega, \Gamma)\Vert p_e(\Omega,\Gamma)\right] \\
    &= \mathbb{E}_{p_g}[-\log p_e(\Omega,\Gamma)],
\end{aligned}
\end{equation}
where $\mathbb{KL}$ denotes the Kullback-Leibler (KL) divergence, $p_g$ denotes the ground-truth distributions\footnote{For simplicity, we abbreviate the probability distribution $p_g(\cdot)$ to $p_g$.} and $p_e$ denotes the estimated distributions. 
Herein, we devise an entropy model to adeptly estimate distributions $p_e(\Omega, \Gamma)$, thus bridging the gap between ground-truth and estimated rate costs.
As depicted in Fig.~\ref{fig::entropy_model}, the inputs of our entropy model comprise reference embeddings $f_\omega$ and covariance $\Sigma_\omega$ of the anchor primitive $\omega$, alongside residual embeddings $f_{\gamma_k}$ from the paired coupled primitive $\gamma_k$.
The entropy model then estimates probability distributions and coding costs for the three parameters.
Specifically, scalar quantization is first applied to the parameters, yielding discrete versions $\{\hat{f}_\omega, \hat{f}_{\gamma_k}, \hat{\Sigma}_\omega\}$.
Then, we seek a differentiable approximation of quantization to enable end-to-end optimization, wherein uniform noise~\cite{balle2018variational} is applied to simulate the effect of scalar quantization. 
The differentiable approximation of $\hat{f}_{\omega}$ can be calculated by: 
\begin{equation}
    \bar{f}_\omega=\frac{f_{\omega}}{\pi_{f_\omega}} + \mathcal{U}(-\frac{\pi_{f_\omega}}{2}, \frac{\pi_{f_\omega}}{2}), 
\end{equation}
where $\pi_{f_\omega}$ denotes the quantization step and $\mathcal{U}$ denotes the uniform noise.
The approximations of $\hat{f}_{\gamma_k}$ and $\hat{\Sigma}_{\omega}$ can be calculated in a similar manner, yielding $\{\bar{f}_{\gamma_k}, \bar{\Sigma}_{\omega}\}$ for subsequent entropy estimation.

In the entropy estimation, context model based on feature grid~\cite{chen2024hac} is harnessed to explore spatial priors $\varrho$, which are then shared among $\{\bar{f}_\omega, \bar{\Sigma}_\omega, \bar{f}_{\gamma_k}\}$.
Additionally, hyperpriors $\eta_\omega$ are extracted from the original signals $f_\omega$ as side information to enrich priors for accurate entropy estimation.
The probability of $\bar{f}_\omega$ is parametrically modeled by a normal distribution, with parameters derived by the spatial priors $\varrho$ and hyperpriors $\eta_\omega$, i.e.,
\begin{equation}
    p_e(\bar{f}_\omega)=\mathcal{N}(\varepsilon_f,\beta_f), \quad \text{with}\; \varepsilon_f,\beta_f = \mathcal{E}_f(\eta_\omega\oplus \varrho),
\end{equation}
where $\{\varepsilon_f,\beta_f\}$ denotes the parameters to determine the normal distribution and $\mathcal{E}_f$ denotes the linear layers for entropy parameter prediction.
The coding costs $R(\bar{f}_\omega)$ is composed of costs of the parameters and the hyperpriors, i.e., 
\begin{equation}
    R(\bar{f}_\omega) = \mathbb{E}_{p_g}\left[-\log p_e(\bar{f}_\omega) -\log p_e(\eta_\omega)\right],
\end{equation}
where $p_e(\eta_\omega)$ denotes the distributions of $\eta_\omega$ estimated by factorized entropy bottleneck~\cite{balle2018variational}.
A similar process applies to $\bar{f}_{\gamma_k}$, yielding the estimated coding costs:
\begin{equation}
    R(\bar{f}_{\gamma_k}) = \mathbb{E}_{p_g}\left[-\log p_e(\bar{f}_{\gamma_k}) -\log p_e(\eta_{\gamma_k})\right],
\end{equation}
where $\eta_{\gamma_k}$ denotes the hyperpriors extracted from $f_{\gamma_k}$.
Moreover, we experimentally observe that hyperpriors cannot substantially benefit the probability estimation of $\bar{\Sigma}_\omega$, and only leverage the spatial contexts $\varrho$ for rate modeling, i.e.,
\begin{equation}
\begin{aligned}
     R(&\bar{\Sigma}_\omega) = \mathbb{E}_{p_g}\left[-\log p_e(\bar{\Sigma}_\omega)\right],\\
      &\text{with}\;\;p_e(\bar{\Sigma}_\omega) = \mathcal{N}(\varepsilon_\Sigma, \beta_\Sigma), \;\varepsilon_\Sigma, \beta_\Sigma = \mathcal{E}_\Sigma(\varrho),
\end{aligned}
\end{equation}
where $\mathcal{E}_\Sigma$ denotes the linear layers for entropy parameter prediction.
Thereafter, the overall rate costs of $\{\Omega, \Gamma\}$ can be formulated as
\begin{equation}
    R(\Omega, \Gamma) = \sum_{\omega \in \Omega} \left[R(\bar{f}_\omega) + R(\bar{\Sigma}_\omega)\right] + \sum_{\gamma_k \in \Gamma} R(\bar{f}_{\gamma_k}).
\end{equation}

Built upon the reliable entropy estimation, we formulate the rate-distortion loss by combining the rendering distortion and the estimated rate. 
Our method is then optimized under this rate-distortion loss to obtain primitives with compact parameters, thereby effectively eliminating the redundancy in Gaussian splatting representations.

\begin{figure}[t]
  \centering
  \includegraphics[width=1\linewidth]{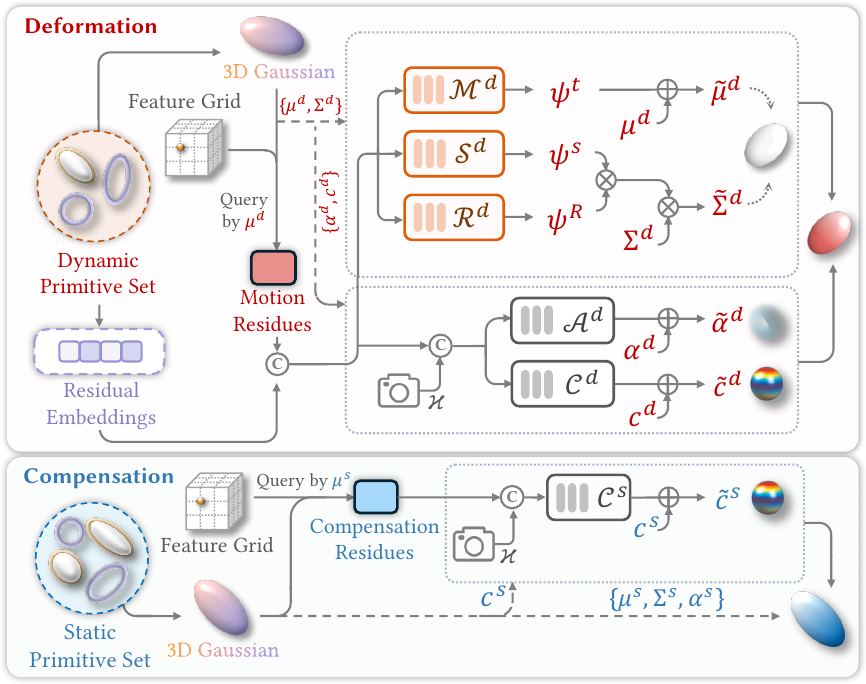}
  \caption{Illustration of the proposed temporal primitive prediction module.}
  \label{fig::temporal_pred}
\end{figure}

\section{Compressed Gaussian Splatting for Dynamic Scenes}
\subsection{Overview}
Fig.~\ref{fig::dynamic_overview} illustrates the extension of our framework to achieve compact representation for dynamic scenes.
As shown in this figure, we leverage the well-established streamable 4DGS paradigm~\cite{sun20243dgstream, gao2024hicom} to model dynamic scenes, and chronologically process $T$ consecutive scenes $B = \{b_1,\ldots,b_T\}$.
The I-scene $b_1$ is compactly modeled using the method proposed in Section~\ref{sec::static}.
For subsequent P-scenes $b_t \in B\backslash \{b_1\}$, we devise a temporal primitive prediction module to compactly construct scene representations, with primitives for $b_t$ adeptly cultivated by referencing their antecedents in $b_{t-1}$.
Specifically, the proposed module capitalizes on primitives $\{\Omega_{t-1}, \Gamma_{t-1}\}$ from the prior scene $b_{t-1}$ to perform prediction.
The process begins by disentangling inherited primitives into a dynamic subset $\{\Omega^d, \Gamma^d\}$ and a static subset $\{\Omega^s, \Gamma^s\}$, where primitive motion confidence serves to guide the classification of these primitives.
Dynamic primitives within $\{\Omega^d, \Gamma^d\}$ necessitate deformation on their geometry and appearance attributes to characterize the forthcoming scene $b_t$.
Static primitives within $\{\Omega^s, \Gamma^s\}$ maintain invariant geometry attributes, with only color attributes undergoing compensation to accommodate ambient lighting changes.
Moreover, to achieve the prediction, we cultivate temporal residues, including motion residues for dynamic primitives and compensation residues for static primitives.
These temporal residues will be further encoded into the bitstream of $b_t$, and serve as priors for the temporal prediction.
With the exploitation of reference primitives, the proposed prediction paradigm can effectively eliminate temporal primitive redundancy and facilitate compact dynamic scene modeling.
The detailed prediction process is further explicated in Section~\ref{sec::temproal_pred}.

Moreover, we propose a temporal adaptive control module, which modifies primitive constitution during optimization to precisely model the emerging scene $b_t$.
Specifically, to mitigate limitations of the initial primitive disentanglement, we devise a mutual conversion mechanism for inherited dynamic and static primitives. 
This mechanism dynamically transforms primitive types based on their behavior during optimization, converting dynamic primitives to static ones and vice versa. 
Such flexible conversion facilitates the efficacy of temporal primitive prediction and enables more accurate scene representation.
Furthermore, we propose to create additional primitives for emerging objects and complex temporal changes when inherited primitives prove insufficient. 
Comprehensive descriptions of the temporal adaptive control mechanism follow in Section~\ref{sec::temporal_ada_ctrl}.

\subsection{Temporal Primitive Prediction}\label{sec::temproal_pred}
As shown in Fig.~\ref{fig::temporal_pred}, our temporal primitive prediction module takes primitives $\{\Omega_{t-1}, \Gamma_{t-1}\}$ inherited from previous scene $b_{t-1}$ as inputs, and outputs 3D Gaussians to adeptly construct the current scene $b_t$.
Specifically, the proposed module involves a primitive disentanglement mechanism, classifying input primitives into dynamic and static sets based on motion confidence.
To derive the motion confidence, we apply motion detection~\cite{xu20244k4d} on $V$ training views of $b_t$ and generate 2D motion masks $E=\{e_1,\ldots, e_V\}$.
For each view $v\in\{1,\ldots, V\}$, we further project 3D Gaussians yielded by $\{\Omega_{t-1}, \Gamma_{t-1}\}$ onto the image plane of $v$, and identify intersections between 2D motion regions of $e_v$ and the projected primitives.
The view-specific motion confidence $\varpi_v$ for primitive pair $\{\omega, \gamma| \omega \in \Omega_{t-1}, \gamma \in \Gamma_{t-1}\}$ can be calculated by:
\begin{equation}
  \varpi_v = \left\{
    \begin{array}{ll}
    1,& \text{if}\; \mathcal{O}(\omega, \gamma)\cap e_v \neq \varnothing,\\
    0,& \text{otherwise},
  \end{array}\right.
\end{equation}
where $\mathcal{O}(\omega, \gamma)$ denotes the projected 3D Gaussian.
We compute the overall motion confidence $\varpi$ of $\{\omega, \gamma\}$ by averaging all view-specific confidences, and then disentangle inherited primitives based on this motion confidence.
Pairs with anchor exhibiting motion confidence below a predefined threshold are classified into the static set $\{\Omega^s, \Gamma^s\}$, while the others form the dynamic set $\{\Omega^d, \Gamma^d\}$.

\begin{table*}
\tiny
\caption{Performance comparison on the Tanks\&Temples dataset~\cite{knapitsch2017tanks}}
\label{tab::t2t}
\centering
\resizebox{0.9\linewidth}{!}{
\begin{tabular}{l|cccr|cccr}
\toprule[0.5pt]
& \multicolumn{4}{c|}{\textit{Train}} & \multicolumn{4}{c}{\textit{Truck}} \\
\cmidrule{2-9} \multicolumn{1}{c|}{}& \multicolumn{1}{c}{PSNR (dB)} & \multicolumn{1}{c}{SSIM} & \multicolumn{1}{c}{LPIPS} & \multicolumn{1}{c|}{Size (MB)} & \multicolumn{1}{c}{PSNR (dB)} & \multicolumn{1}{c}{SSIM} & \multicolumn{1}{c}{LPIPS} & \multicolumn{1}{c}{Size (MB)}\\
\midrule[0.3pt]
Kerbl et~al.~\cite{kerbl20233d}                   & 22.02 & 0.81 & 0.21 & 257.44 & 25.41 & 0.88 & 0.15 & 611.31 \\
Navaneet et~al.~\cite{navaneet2025compgs}         & 21.63 & 0.80 & 0.22 &  27.54 & 25.04 & 0.88 & 0.16 &  66.48 \\
Niedermayr et~al.~\cite{niedermayr2024compressed} & 21.92 & 0.81 & 0.22 &  13.78 & 25.24 & 0.88 & 0.15 &  21.51 \\
Lee et~al.~\cite{lee2024compact}                  & 21.69 & 0.80 & 0.24 &  37.38 & 25.10 & 0.87 & 0.16 &  41.55 \\
Girish et~al.~\cite{girish2024eagles}             & 21.68 & 0.80 & 0.23 &  24.67 & 25.10 & 0.87 & 0.17 &  42.46 \\
Liu et~al.~\cite{liu2024compgs}-\textit{high}     & 22.12 & 0.80 & 0.23 &   8.60 & 25.28 & 0.87 & 0.18 &  10.61 \\
Liu et~al.~\cite{liu2024compgs}-\textit{middle}   & 21.82 & 0.80 & 0.24 &   6.72 & 24.97 & 0.86 & 0.20 &   7.82 \\
Liu et~al.~\cite{liu2024compgs}-\textit{low}      & 21.49 & 0.78 & 0.26 &   5.51 & 24.72 & 0.85 & 0.21 &   6.27 \\
Chen et~al.~\cite{chen2024hac}-\textit{high}      & 22.53 & 0.83 & 0.20 &  11.21 & 26.01 & 0.89 & 0.14 &  15.31 \\
Chen et~al.~\cite{chen2024hac}-\textit{middle}    & 22.42 & 0.82 & 0.21 &   8.58 & 25.94 & 0.88 & 0.15 &  10.09 \\
Chen et~al.~\cite{chen2024hac}-\textit{low}       & 22.37 & 0.82 & 0.21 &   7.09 & 25.87 & 0.88 & 0.16 &   9.28 \\
\midrule[0.3pt]
Proposed-\textit{high}                            & 22.60 & 0.82 & 0.22 &   5.51 & 25.45 & 0.87 & 0.18 &   5.06 \\
Proposed-\textit{middle}                          & 22.56 & 0.81 & 0.22 &   4.92 & 25.33 & 0.87 & 0.19 &   4.00 \\
Proposed-\textit{low}                             & 22.44 & 0.81 & 0.23 &   3.91 & 25.09 & 0.86 & 0.20 &   3.09 \\
\bottomrule[0.5pt]
\end{tabular}}
\end{table*}

Deformation is then applied to dynamic primitives and yields 3D Gaussians aligned to the current scene $b_t$.
In this process, we cultivate motion residues for each dynamic primitive pair $\{\omega^d, \gamma^d\}$ to generate deformation fields, followed by the geometry warping on the 3D Gaussian yielded by $\{\omega^d, \gamma^d\}$.
Specifically, we leverage feature grid~\cite{chen2024hac} to compactly store the motion residues $\rho^d$ and use locations of the original 3D Gaussian yielded by $\{\Omega_{t-1}, \Gamma_{t-1}\}$ to retrieve $\rho^d$.
Meanwhile, residual embeddings $f_{\gamma}$ of $\gamma^d$, which contains initial spatial residual information, are fused with $\rho^d$ to generate representative prediction features $\zeta^d$.
The deformation fields on moment $t$ are further predicted based on $\zeta^d$, i.e.,
\begin{equation}
  \psi^t = \mathcal{M}^d(\zeta^d), \quad \psi^s = \mathcal{S}^d(\zeta^d), \quad \psi^R = \mathcal{R}^d(\zeta^d),
\end{equation}
where $\{\mathcal{M}^d, \mathcal{S}^d, \mathcal{R}^d\}$ denote prediction linear layers and $\{\psi^t, \psi^s, \psi^R\}$ denote the translation, scaling, and rotation components within the deformation field, respectively.
Subsequently, the geometry attributes of the predicted 3D Gaussian are generated via affine transform, i.e.,
\begin{equation}
  \tilde{\mu}^d = \mu^d + \psi^t, \quad \tilde{\Sigma}^d = Diag(\psi^s)\times \psi^R \times \Sigma^d,
\end{equation}
where $\{\mu^d, \Sigma^d\}$ denote geometry attributes of the original 3D Gaussian yielded by $\{\omega^d, \gamma^d\}$.
Simultaneously, appearance offsets are predicted based on the view embeddings $\varkappa$ and $\zeta^d$, compensating the appearance attributes of the original 3D Gaussian, i.e.,
\begin{equation}
  \tilde{c}^d = c^d + \mathcal{C}^d(\varkappa \oplus \zeta^d), \quad \tilde{\alpha}^d = \alpha^d + \mathcal{A}^d(\varkappa \oplus \zeta^d),
\end{equation}
where $\{\mathcal{C}^d, \mathcal{A}^d\}$ denote the prediction linear layers and $\{c^d, \alpha^d\}$ denote the appearance attributes of the original 3D Gaussian.
Furthermore, the remaining static primitives $\{\Omega^s, \Gamma^s\}$ are calibrated to accommodate the shift of global illumination, with appearance attributes adjusted and geometry attributes preserved.
To achieve this, we devise compensation residues $\rho^s$ amending appearance adjustment, with $\rho^s$ represented via feature grid~\cite{chen2024hac} for compact storage.
We experimentally observe that adjusting opacity exerts negligible benefits to the prediction, and only predict the color offsets based on $\rho^s$ and the view embeddings $\varkappa$, i.e.,
\begin{equation}
  \tilde{c}^s = c^s + \mathcal{C}^s(\varkappa \oplus \rho^s),
\end{equation}
where $\mathcal{C}^s$ denotes the prediction linear layers, $c^s$ and $\tilde{c}^s$ denote the color of the original and the compensated 3D Gaussians, respectively.

With the devised temporal prediction, the scene $b_t$ can be efficiently represented by inherited primitives and succinct temporal residues, thereby achieving compact dynamic scene modeling.

\subsection{Temporal Adaptive Control}\label{sec::temporal_ada_ctrl}
The inherent imperfections in 2D motion detection necessitate effective refinement to address primitive misclassification.
Hence, we propose a temporal adaptive control module to facilitate the management of primitives, including the mutual conversion between dynamic and static primitives and the creation of new primitives when required.
Specifically, primitives related to moving objects might be incorrectly identified as static in initialization.
Consequently, compensation for color attributes alone proves inadequate and leads to substantial rendering errors. 
To address this issue, we use the positional gradients~\cite{kerbl20233d} of 3D Gaussians yielded by static primitives as a surrogate measure for rendering error, and ascertain whether static primitives require a dynamic status transition. 
The static-to-dynamic conversion can be represented as:
\begin{equation}
\begin{aligned}
  \{\Omega^{s\rightarrow d}, \Gamma^{s\rightarrow d}\} =& \\
 \big\{\omega^s, \gamma^s| \{\omega^s, &\gamma^s\} \in \{\Omega^s, \Gamma^s\},\triangledown_g({\omega^s}) > \tau_{s\rightarrow d}\big\},
\end{aligned}
\end{equation}
where $\triangledown_g({\omega^s})$ denotes the accumulated positional gradients~\cite{kerbl20233d} for all 3D Gaussians yielded by $\omega^s$, and $\tau_{s\rightarrow d}$ represents the threshold value.
Meanwhile, primitives affiliated with stationary objects may be misclassified as dynamic and produce geometry artifacts.
Therefore, we further propose a dynamic-to-static conversion mechanism, mitigating fake geometry compensation by reclassifying dynamic primitives with insignificant geometry offsets.
Specifically, for each 3D Gaussian derived by primitive pairs within $\{\Omega^d, \Gamma^d\}$, we comprehensively assess the significance of geometry deformations through a compositional measure of translation, scaling, and rotation. 
The significance of translation and scaling is evaluated by the L1 norm of the respective offsets, while rotation significance is quantified through cosine similarity between the rotation quaternion and the identity quaternion. 
The overall significance of the deformed 3D Gaussian is formulated as:
\begin{equation}
  \vartheta =  \|\psi^t\|_1 + \|\psi^s\|_1 + 1 - \frac{q(\psi^R) \cdot q_i}{\|q(\psi^R)\|_2 \|q_i\|_2},
\end{equation}
where $q(\psi^R)$ denotes quaternions derived from the rotation matrix $\psi^R$, and $q_i$ denotes the identity quaternion.
Accordingly, the dynamic-to-static conversion process is conducted under threshold $\tau_{d\rightarrow s}$, i.e., 
\begin{equation}
\begin{aligned}
  \{\Omega^{d\rightarrow s}, \Gamma^{d\rightarrow s}\} =& \\
  \big\{\omega^d, \gamma^d| \{\omega^d, &\gamma^d\} \in \{\Omega^d, \Gamma^d\},\vartheta(\omega^d) < \tau_{d\rightarrow s}\big\},
\end{aligned}
\end{equation}
where $\vartheta(\omega^d)$ denotes the accumulated significance for all deformed 3D Gaussians yielded by $\omega^d$.

Furthermore, inherited primitives may be insufficient to model emerging objects and drastic movements within $b_t$.
Hence, we devise a primitive creation mechanism to add necessary primitives during the optimization.
This creation process is driven by positional gradients~\cite{kerbl20233d}, with new primitives introduced in regions exhibiting high gradient values. 
Importantly, we set the creation threshold higher than the threshold that triggers static-to-dynamic conversion.
This design prioritizes the use of dynamic primitives when capturing motion changes, leveraging their superior coding efficiency.

\section{Experiments}

\begin{table*}
\tiny
\caption{Performance comparison on the Deep Blending dataset~\cite{hedman2018deep}}
\label{tab::db}
\centering
\resizebox{0.9\linewidth}{!}{
\begin{tabular}{l|cccr|cccr}
\toprule[0.5pt]
& \multicolumn{4}{c|}{\textit{DrJohnson}} & \multicolumn{4}{c}{\textit{Playroom}} \\
\cmidrule{2-9} \multicolumn{1}{c|}{}& \multicolumn{1}{c}{PSNR (dB)} & \multicolumn{1}{c}{SSIM} & \multicolumn{1}{c}{LPIPS} & \multicolumn{1}{c|}{Size (MB)} & \multicolumn{1}{c}{PSNR (dB)} & \multicolumn{1}{c}{SSIM} & \multicolumn{1}{c}{LPIPS} & \multicolumn{1}{c}{Size (MB)}\\
\midrule[0.3pt]
Kerbl et~al.~\cite{kerbl20233d} & 29.14 & 0.90 & 0.24 & 782.10 & 29.94 & 0.91 & 0.24 & 549.88 \\
Navaneet et~al.~\cite{navaneet2025compgs}  & 29.34 & 0.90 & 0.25 & 85.10 & 30.43 & 0.91 & 0.24 & 59.82 \\
Niedermayr et~al.~\cite{niedermayr2024compressed} & 29.03 & 0.90 & 0.25 & 27.86 & 29.86 & 0.91 & 0.25 & 19.88 \\
Lee et~al.~\cite{lee2024compact}  & 29.26 & 0.90 & 0.25 & 48.11 & 30.38 & 0.91 & 0.25 & 38.17 \\
Girish et~al.~\cite{girish2024eagles} & 29.52 & 0.91 & 0.24 & 80.09 & 30.27 & 0.91 & 0.25 & 43.28 \\
Liu et~al.~\cite{liu2024compgs}-\textit{high} & 29.33 & 0.90 & 0.27 & 10.38 & 30.04 & 0.90 & 0.29 & 7.15 \\
Liu et~al.~\cite{liu2024compgs}-\textit{middle} & 29.21 & 0.90 & 0.27 & 8.21 & 29.59 & 0.89 & 0.30 & 5.43 \\
Liu et~al.~\cite{liu2024compgs}-\textit{low} & 28.99 & 0.90 & 0.28 & 7.00 &  29.61 & 0.89 & 0.31 & 5.06 \\
Chen et~al.~\cite{chen2024hac}-\textit{high} & 29.77  & 0.91  & 0.25  & 8.68 & 30.76  & 0.91  & 0.26  & 6.27 \\
Chen et~al.~\cite{chen2024hac}-\textit{middle} & 29.70 & 0.91 & 0.26 & 7.94 & 30.68  & 0.91  & 0.26  & 4.17 \\
Chen et~al.~\cite{chen2024hac}-\textit{low} & 29.55 & 0.91 & 0.26 & 5.72 & 30.41  & 0.91  & 0.27  & 3.23 \\
\midrule[0.3pt]
Proposed-\textit{high} & 29.76 & 0.91 & 0.26 & 5.86 & 30.80 & 0.91 & 0.27 & 4.50 \\
Proposed-\textit{middle} & 29.50 & 0.90 & 0.27 & 3.88 & 30.52 & 0.90 & 0.28 & 2.71 \\
Proposed-\textit{low} & 28.88 & 0.89 & 0.29 & 2.12 & 29.59 & 0.89 & 0.30 & 1.48 \\
\bottomrule[0.5pt]
\end{tabular}}
\end{table*}

\subsection{Implementation Details}
The dimensions of reference and residual embeddings are set to 32 and 4, respectively.
The ratio $K$ of coupled primitives to anchor primitives is experimentally determined as 10, and the quantization steps are adaptively predicted using the corresponding prior features. 
Meanwhile, linear layer configurations follow our previous implementation~\cite{liu2024compgs}, while feature grids are configured according to~\cite{chen2024hac}.
To obtain multiple bitrate points, we set the Lagrange multiplier $\lambda$ to $\{0.0001, 0.0005, 0.001\}$. 
For the final compression stage, we employ the G-PCC point cloud codec~\cite{schwarz2018emerging} to compress anchor primitive locations, while the remaining parameters are compressed using the arithmetic coding library torchAC~\cite{mentzer2019practical} based on the estimated distributions.

\subsection{Experiment Evaluation on Static Scenes}
\textbf{Experimental settings.}
We conduct extensive evaluations of our method across prevailing benchmark datasets~\cite{knapitsch2017tanks, hedman2018deep, barron2022mip}, and strictly adhere to the experimental protocol established by 3DGS~\cite{kerbl20233d} to ensure fair comparisons.
Specifically, we evaluate our method on the scenes utilized by 3DGS~\cite{kerbl20233d}: the \textit{Train} and \textit{Truck} scenes from Tanks\&Temples~\cite{knapitsch2017tanks}, the \textit{DrJohnson} and \textit{Playroom} scenes from Deep Blending~\cite{hedman2018deep}, and both indoor and outdoor scenes from Mip-NeRF 360~\cite{barron2022mip}.
We maintain consistency with 3DGS~\cite{kerbl20233d} regarding training and test view splits, camera poses, and initialized point clouds.
For quantitative assessment, we employ PSNR, SSIM~\cite{wang2004image}, and LPIPS~\cite{zhang2018unreasonable} metrics to evaluate rendering quality, while using model size as the measure of coding efficiency. 

We establish 3DGS~\cite{kerbl20233d} as the uncompressed baseline and include comprehensive comparisons with state-of-the-art Gaussian splatting compression methods~\cite{navaneet2025compgs, niedermayr2024compressed, lee2024compact, girish2024eagles, liu2024compgs, chen2024hac}. 
These comparative methods are retrained using their default configurations. 
For methods supporting multiple bitrate configurations~\cite{liu2024compgs, chen2024hac}, we select three bitrate points for comparison.

\begin{table*}[t]
\tiny
\centering
\caption{Performance comparison on the Mip-NeRF 360 dataset~\cite{barron2022mip}}
\label{tab::mip}
\resizebox{0.9\linewidth}{!}{
\begin{tabular}{l|cccr|cccr}
\toprule[0.5pt]
& \multicolumn{4}{c|}{\textit{Outdoor}} & \multicolumn{4}{c}{\textit{Indoor}} \\
\cmidrule{2-9} \multicolumn{1}{c|}{}& \multicolumn{1}{c}{PSNR (dB)} & \multicolumn{1}{c}{SSIM} & \multicolumn{1}{c}{LPIPS} & \multicolumn{1}{c|}{Size (MB)} & \multicolumn{1}{c}{PSNR (dB)} & \multicolumn{1}{c}{SSIM} & \multicolumn{1}{c}{LPIPS} & \multicolumn{1}{c}{Size (MB)}\\
\midrule[0.3pt]
Kerbl et~al.~\cite{kerbl20233d}                   & 24.63 & 0.73 & 0.24 & 969.38 & 30.98 & 0.93 & 0.19 & 344.00 \\
Navaneet et~al.~\cite{navaneet2025compgs}         & 24.47 & 0.72 & 0.25 & 111.03 & 30.26 & 0.92 & 0.20 &  36.74 \\
Niedermayr et~al.~\cite{niedermayr2024compressed} & 24.40 & 0.71 & 0.26 &  37.77 & 30.52 & 0.92 & 0.20 &  14.95 \\
Lee et~al.~\cite{lee2024compact}                  & 24.33 & 0.71 & 0.27 &  51.98 & 30.45 & 0.92 & 0.20 &  37.38 \\
Girish et~al.~\cite{girish2024eagles}             & 24.34 & 0.71 & 0.27 &  74.24 & 30.41 & 0.92 & 0.20 &  40.18 \\
Liu et~al.~\cite{liu2024compgs}-\textit{high}     & 24.56 & 0.71 & 0.27 &  22.48 & 30.64 & 0.92 & 0.21 &  10.16 \\
Liu et~al.~\cite{liu2024compgs}-\textit{middle}   & 24.35 & 0.70 & 0.29 &  16.53 & 29.83 & 0.90 & 0.22 &   7.06 \\
Liu et~al.~\cite{liu2024compgs}-\textit{low}      & 24.13 & 0.69 & 0.31 &  14.25 & 29.17 & 0.89 & 0.24 &   5.53 \\
Chen et~al.~\cite{chen2024hac}-\textit{high}      & 24.76 & 0.72 & 0.25 &  35.45 & 31.52 & 0.93 & 0.19 &  19.90 \\
Chen et~al.~\cite{chen2024hac}-\textit{middle}    & 24.63 & 0.72 & 0.26 &  28.55 & 31.26 & 0.92 & 0.20 &  16.02 \\
Chen et~al.~\cite{chen2024hac}-\textit{low}       & 24.64 & 0.72 & 0.26 &  24.95 & 30.91 & 0.92 & 0.20 &  13.10 \\
\midrule[0.3pt]
Proposed-\textit{high}                            & 24.60 & 0.71 & 0.28 &  18.32 & 31.27 & 0.92 & 0.20 &   6.81 \\
Proposed-\textit{middle}                          & 24.53 & 0.71 & 0.29 &  15.00 & 30.85 & 0.91 & 0.21 &   4.92 \\
Proposed-\textit{low}                             & 24.37 & 0.70 & 0.30 &  12.06 & 30.43 & 0.91 & 0.22 &   3.63 \\
\bottomrule[0.5pt]
\end{tabular}}
\end{table*}

\begin{figure*}
\centering
\includegraphics[width=1.\linewidth]{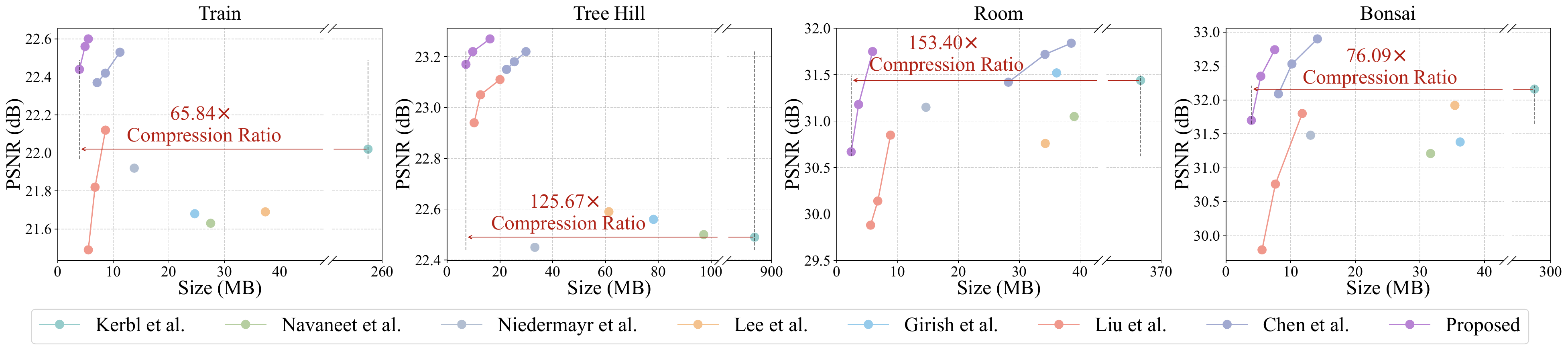}
\caption{Rate-distortion curves of the proposed method and existing works~\cite{navaneet2025compgs, niedermayr2024compressed, lee2024compact, girish2024eagles, liu2024compgs, chen2024hac} on several representative scene.}
\label{fig::rd_curves}
\end{figure*}

\textbf{Quantitative results.}
Quantitative results presented in Table~\ref{tab::t2t} demonstrate that the proposed method achieves remarkable performance on the Tanks\&Temples dataset~\cite{knapitsch2017tanks}.
For the \textit{Train} scene, our method reduces the model size from 257.44~ MB to a minimum of 3.91~MB when compared to the uncompressed 3DGS~\cite{kerbl20233d}, achieving a significant compression ratio of 65.84$\times$.
Meanwhile, the proposed method exhibits a significant advantage over existing compression methods~\cite{navaneet2025compgs, niedermayr2024compressed, lee2024compact, girish2024eagles, liu2024compgs, chen2024hac}.
Notably, our method achieves more than 50\% size reduction compared to the state-of-the-art method~\cite{chen2024hac} while maintaining comparable rendering quality around 22.56~dB. 
These improvements stem from our spatial primitive prediction and rate-constrained optimization modules, which effectively eliminate redundancies within Gaussian splatting representations.
Results on the \textit{Truck} scene further exemplify the effectiveness of the proposed method. Compared to 3DGS~\cite{kerbl20233d}, our method reduces the model size from 611.31~MB to a minimum of~3.09 MB while maintaining competitive rendering quality, attaining an impressive compression ratio of 197.83$\times$.
Table~\ref{tab::db} presents comparison results on the Deep Blending dataset~\cite{hedman2018deep}. 
The proposed method demonstrates remarkable compression effectiveness compared to 3DGS~\cite{kerbl20233d}, reducing the model size from 782.10~MB to 2.12~MB for the \textit{DrJohnson} scene and from 549.88~MB to 1.48~MB for the \textit{Playroom} scene.
Meanwhile, our method surpasses all comparative methods~\cite{navaneet2025compgs, niedermayr2024compressed, lee2024compact, girish2024eagles, liu2024compgs, chen2024hac}, cultivating impressive compression ratios of 368.92$\times$ and 371.54$\times$ on the two scenes, respectively.
Moreover, results on the Mip-NeRF 360 dataset~\cite{barron2022mip} elaborate the superiority of the proposed method, as depicted in Table~\ref{tab::mip}. 
For outdoor scenes, the proposed method achieves an average compression ratio of 80.36$\times$ compared to 3DGS~\cite{kerbl20233d}, while reaching an average of 94.83$\times$ for indoor scenes. 
Our method presents significant advancements over existing works~\cite{navaneet2025compgs, niedermayr2024compressed, lee2024compact, girish2024eagles, liu2024compgs, chen2024hac}, where these methods achieve compression ratios between 8.73$\times$ and 68.03$\times$ for outdoor scenes and between 8.56$\times$ and 69.99$\times$ for indoor scenes.
Furthermore, we provide rate-distortion curves on several representative scenes in Fig.~\ref{fig::rd_curves} to intuitively demonstrate the superiority of our method.
It can be observed that, compared to extant methods~\cite{navaneet2025compgs, niedermayr2024compressed, lee2024compact, girish2024eagles, liu2024compgs, chen2024hac}, the proposed method achieves substantial improvements in both rendering quality and representation compactness.

\begin{figure*}
\centering
\includegraphics[width=0.85\linewidth]{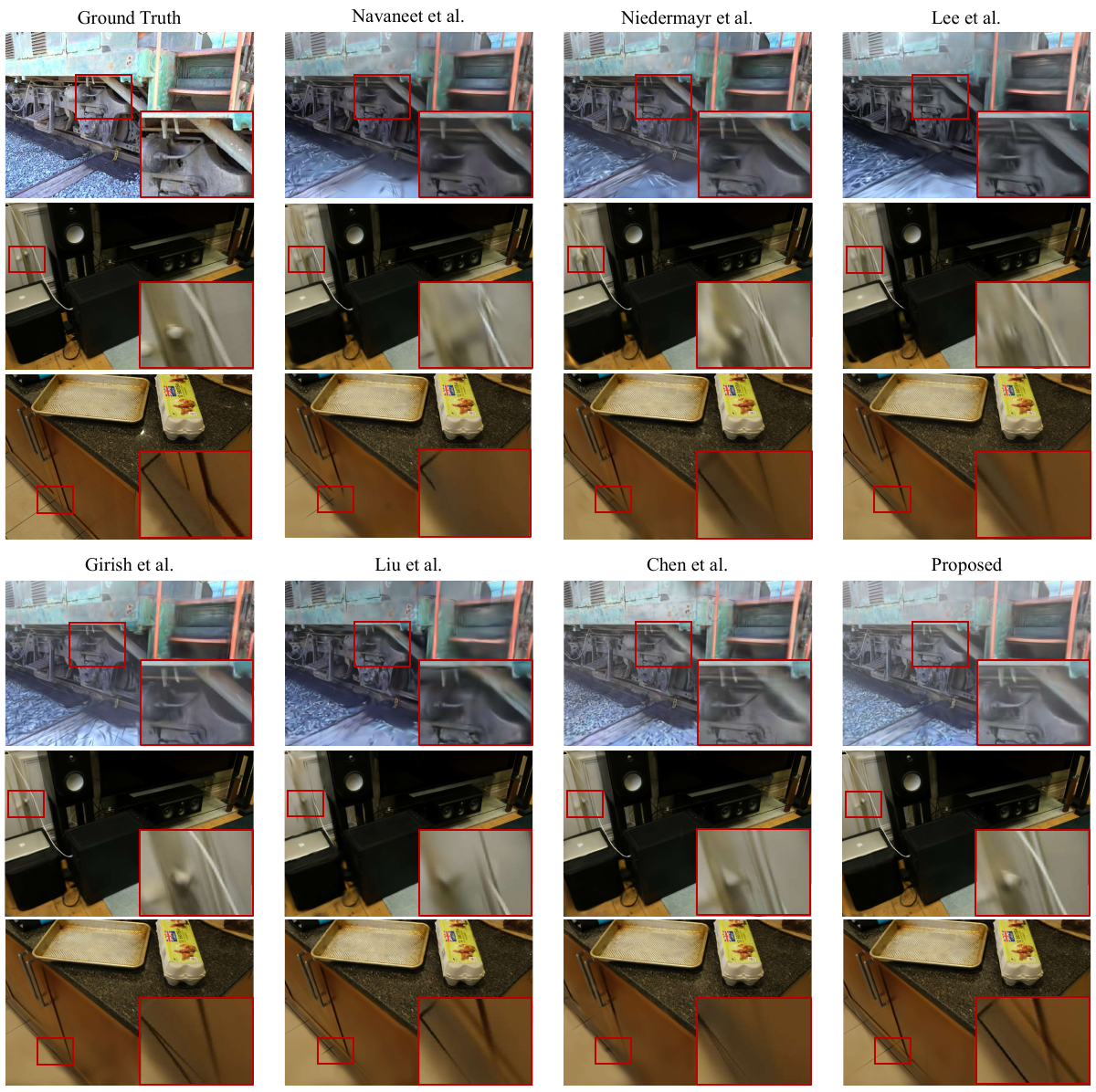}
\caption{Qualitative comparison between the proposed method and existing works~\cite{navaneet2025compgs, niedermayr2024compressed, lee2024compact, girish2024eagles, liu2024compgs, chen2024hac}.}
\label{fig::visual_static}
\end{figure*}

\textbf{Qualitative results.}
Fig.~\ref{fig::visual_static}  provides qualitative comparisons between the proposed method and prior compression methods~\cite{navaneet2025compgs, niedermayr2024compressed, lee2024compact, girish2024eagles, liu2024compgs, chen2024hac}, with zoomed-in details for visualization.  
The results demonstrate that the proposed method can capture intricate scene details and yield rendered views of superior visual quality.

\subsection{Experiment Results on Dynamic Scenes}

\textbf{Experimental settings.}
We evaluate our method on the Neural 3D Video dataset~\cite{li2022neural}.
This dataset encompasses diverse scenes featured with intricate details and complex motions, providing a comprehensive benchmark for dynamic modeling. 
Following the experimental protocol of HiCoM~\cite{gao2024hicom}, we utilize the first 300 frames of each scene in evaluation, with the middle view reserved for testing and the remaining views used for training.
Compared methods encompasses both streamable~\cite{sun20243dgstream, gao2024hicom} and non-streamable~\cite{wu20244d, yang2023real, luiten2024dynamic} works. 
Performance evaluation incorporates two key items: PSNR values of the test view to assess rendering quality, and overall model size to measure representation compactness. 
For streamable methods~\cite{sun20243dgstream, gao2024hicom}, we report per-frame model size to facilitate a direct comparison of compression efficiency.
Meanwhile, results of methods~\cite{sun20243dgstream, gao2024hicom} are adopted from HiCoM~\cite{gao2024hicom} since our experimental setup rigorously mirrors their method.

\begin{table*}[t]
\centering
\begin{threeparttable}[b]
\caption{Performance comparison on the Neural 3D Video dataset~\cite{li2022neural}}
\label{tab::neural3dv}
\begin{tabular}{l|crr|crr}
\toprule
& \multicolumn{3}{c|}{\textit{Coffee Martini}} & \multicolumn{3}{c}{\textit{Cook Spinach}} \\
\cmidrule{2-7} \multicolumn{1}{c|}{} & \multicolumn{1}{c}{PSNR (dB)} & \multicolumn{1}{c}{Total Size (MB)} & \multicolumn{1}{c|}{Per Frame Size (MB)} &\multicolumn{1}{c}{PSNR (dB)} & \multicolumn{1}{c}{Total Size (MB)} & \multicolumn{1}{c}{Per Frame Size (MB)}\\
\midrule
Wu et. al.~\cite{wu20244d} & 28.39 & 39.42 & - & 32.61 & 41.59 & - \\
Yang et. al.~\cite{yang2023real} & 27.98 & 3704.58 & - & 32.73 & 2474.94 & - \\
Luiten et. al.~\cite{luiten2024dynamic} & 26.49 & 2760.0 & - & 32.97 & 2760.0 & - \\
Sun et. al.~\cite{sun20243dgstream}$^{\dagger}$ & 26.73 & 2280.0 & 7.6 & 31.38 & 2280.0 & 7.6 \\
Gao et. al.~\cite{gao2024hicom}$^{\dagger}$ & 28.19 & 240.0 & 0.8 & 32.34 & 180.0 & 0.6 \\
\midrule
Proposed-\textit{high}$^{\dagger}$ & 28.68 & 15.31 & 0.05 & 32.48 & 21.50 & 0.07 \\
Proposed-\textit{middle}$^{\dagger}$ & 28.58 & 14.66 & 0.05 & 32.38 & 18.88 & 0.06 \\
Proposed-\textit{low}$^{\dagger}$ & 28.37 & 13.79 & 0.05 & 32.13 & 16.55 & 0.06 \\
\midrule
& \multicolumn{3}{c|}{\textit{Cut Roasted Beef}} & \multicolumn{3}{c}{\textit{Flame Salmon}} \\
\cmidrule{2-7} \multicolumn{1}{c|}{} & \multicolumn{1}{c}{PSNR (dB)} & \multicolumn{1}{c}{Total Size (MB)} & \multicolumn{1}{c|}{Per Frame Size (MB)} &\multicolumn{1}{c}{PSNR (dB)} & \multicolumn{1}{c}{Total Size (MB)} & \multicolumn{1}{c}{Per Frame Size (MB)}\\
\midrule
Wu et. al.~\cite{wu20244d} & 30.20 & 40.90 & - & 29.36 & 41.57 & - \\
Yang et. al.~\cite{yang2023real} & 33.23 & 2555.56 & - & 28.86 & 4695.46 & - \\
Luiten et. al.~\cite{luiten2024dynamic} & 30.72 & 2760.0 & - & 26.92 & 2760.0 & - \\
Sun et. al.~\cite{sun20243dgstream}$^{\dagger}$ & 31.36 & 2280.0 & 7.6 & 27.45 & 2280.0 & 7.6 \\
Gao et. al.~\cite{gao2024hicom}$^{\dagger}$ & 32.36 & 180.0 & 0.6 & 28.54 & 270.0 & 0.9 \\
\midrule
Proposed-\textit{high}$^{\dagger}$ & 32.34 & 23.30 & 0.08 & 28.95 & 17.09 & 0.06 \\
Proposed-\textit{middle}$^{\dagger}$ & 32.40 & 20.20 & 0.07 & 28.85 & 15.89 & 0.05 \\
Proposed-\textit{low}$^{\dagger}$ & 31.96 & 19.97 & 0.07 & 28.53 & 14.79 & 0.05 \\
\midrule
& \multicolumn{3}{c|}{\textit{Flame Steak}} & \multicolumn{3}{c}{\textit{Sear Steak}} \\
\cmidrule{2-7} \multicolumn{1}{c|}{} & \multicolumn{1}{c}{PSNR (dB)} & \multicolumn{1}{c}{Total Size (MB)} & \multicolumn{1}{c|}{Per Frame Size (MB)} &\multicolumn{1}{c}{PSNR (dB)} & \multicolumn{1}{c}{Total Size (MB)} & \multicolumn{1}{c}{Per Frame Size (MB)}\\
\midrule
Wu et. al.~\cite{wu20244d} & 32.96 & 38.81 & - & 33.27 & 39.46 & - \\
Yang et. al.~\cite{yang2023real} & 33.19 & 3173.37 & - & 33.44 & 2164.07 & - \\
Luiten et. al.~\cite{luiten2024dynamic} & 33.24 & 2760.0 & - & 33.68 & 2760.0 & - \\
Sun et. al.~\cite{sun20243dgstream}$^{\dagger}$ & 31.56 & 2280.0 & 7.6 & 32.44 & 2280.0 & 7.6 \\
Gao et. al.~\cite{gao2024hicom}$^{\dagger}$ & 32.13 & 180.0 & 0.6 & 32.42 & 180.0 & 0.6 \\
\midrule
Proposed-\textit{high}$^{\dagger}$ & 33.95 & 22.44 & 0.07 & 33.33 & 18.61 & 0.06 \\
Proposed-\textit{middle}$^{\dagger}$ & 33.86 & 20.21 & 0.07 & 33.22 & 17.50 & 0.06 \\
Proposed-\textit{low}$^{\dagger}$ & 33.57 & 18.39 & 0.06 & 32.95 & 16.39 & 0.05 \\
\bottomrule
\end{tabular}
\begin{tablenotes}
  \item The superscript $\dagger$ denotes the streamable method.
\end{tablenotes}
\end{threeparttable}
\end{table*}
  
\begin{figure*}[t]
  \centering
  \includegraphics[width=0.9\linewidth]{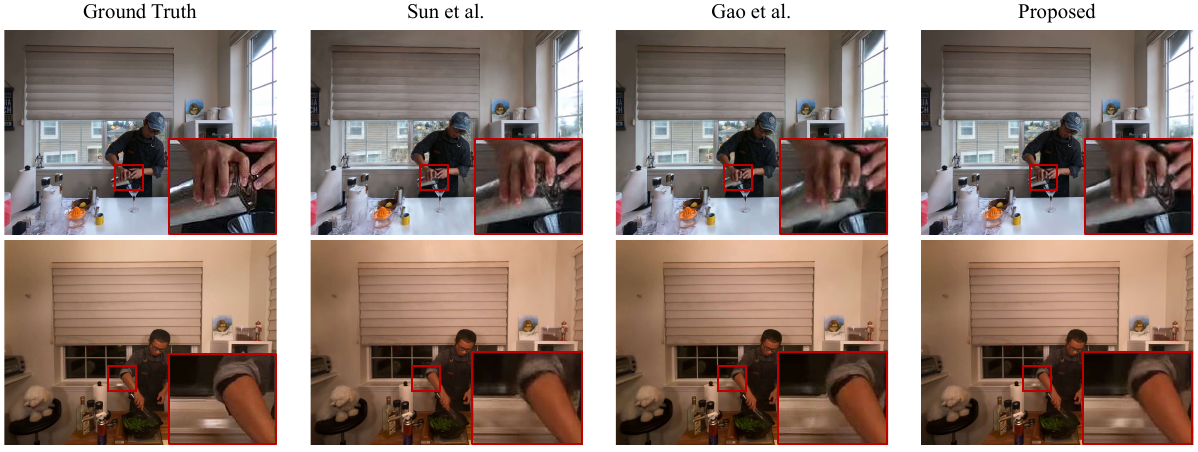}
  \caption{Qualitative comparison between the proposed method and existing 4DGS works~\cite{sun20243dgstream, gao2024hicom}.}
  \label{fig::visual_n3dv}
\end{figure*}  

\textbf{Quantitative results.}
Table~\ref{tab::neural3dv} presents comparison results on the Neural 3D Video dataset~\cite{li2022neural}. 
It can be observed that the proposed method demonstrates significant advancements in both scene representation quality and compactness when compared to existing streamable methods~\cite{sun20243dgstream, gao2024hicom}.
Results on the \textit{Coffee Martini} scene demonstrate these improvements, in which our method achieves a rendering quality of 28.37~dB while maintaining a remarkably efficient rate consumption of just 0.05~MB per frame.
For comparison, 3DGStream~\cite{sun20243dgstream} requires 7.6~MB per frame while achieving a lower rendering quality of 26.73~dB.
HiCoM~\cite{gao2024hicom} achieves 28.19~dB but requires 0.8~MB per frame.
Consistent advancements are observed in the remaining scenes, where the proposed method outperforms the compared methods~\cite{sun20243dgstream, gao2024hicom} in both rendering quality and representation compactness.
The superior performance of our method stems from the innovative temporal compact modeling paradigm.
This paradigm efficiently leverages reference information from previous scenes to predict subsequent scenes, and necessitates only minimal temporal residues to be encoded, thereby enabling highly compact dynamic scene representations.
Moreover, our method achieves performance comparable to non-streamable approaches~\cite{wu20244d, yang2023real, luiten2024dynamic}, despite operating without access to future timestamps. 
This remarkable outcome represents a significant advancement of our method in practical 3D reconstruction applications.

\begin{table*}
\caption{Ablation studies on the Tanks\&Temples dataset~\cite{knapitsch2017tanks}}
\label{tab::ablation_studies}
\centering
\begin{tabular}{cc|cccrcccr}
\toprule
\multicolumn{1}{c}{Spatial Primitive} & \multicolumn{1}{c|}{Rate-constrained}  & \multicolumn{4}{c}{\textit{Train}}  & \multicolumn{4}{c}{\textit{Truck}}  \\
\cmidrule{3-10} \multicolumn{1}{c}{Prediction}  & \multicolumn{1}{c|}{Optimization}  & \multicolumn{1}{c}{PSNR (dB)} & \multicolumn{1}{c}{SSIM} & \multicolumn{1}{c}{LPIPS} & \multicolumn{1}{c}{Size (MB)} & \multicolumn{1}{c}{PSNR (dB)} & \multicolumn{1}{c}{SSIM} & \multicolumn{1}{c}{LPIPS} & \multicolumn{1}{c}{Size (MB)}  \\
\midrule
$\times$      & $\times$      & 22.02  & 0.81  & 0.21  & 257.44  & 25.41  & 0.88  & 0.15  & 611.31  \\
$\checkmark$  & $\times$      & 22.43  & 0.82  & 0.22  &  79.88  & 25.14  & 0.86  & 0.19  &  47.34  \\
$\checkmark$  & $\checkmark$  & 22.44  & 0.81  & 0.23  &  3.91  & 25.09  & 0.86  & 0.20  &  3.09  \\
\bottomrule
\end{tabular}
\end{table*}

\textbf{Qualitative results.}
Fig.~\ref{fig::visual_n3dv} shows qualitative comparisons between the proposed method and existing streamable methods~\cite{sun20243dgstream, gao2024hicom}.
The views rendered by the proposed method showcase smoother boundaries and intricate details within regions of motion, underscoring the effectiveness of our dynamic modeling paradigm. 
Compared to existing works, our method achieves superior quality in background regions, such as the windowsill boundaries. 
This improvement stems from a fundamental difference in modeling strategy. 
Previous methods~\cite{sun20243dgstream, gao2024hicom} indiscriminately apply motion offsets to all primitives and lead to artifacts in static regions where motion modeling is unnecessary. 
In contrast, our method employs a sophisticated primitive disentanglement strategy that differentiates between dynamic and static primitives, which prevents spurious motion offsets in static backgrounds and results in more precise modeling.

\subsection{Ablation Studies}
\begin{figure}[t]
  \centering
  \includegraphics[width=1\linewidth]{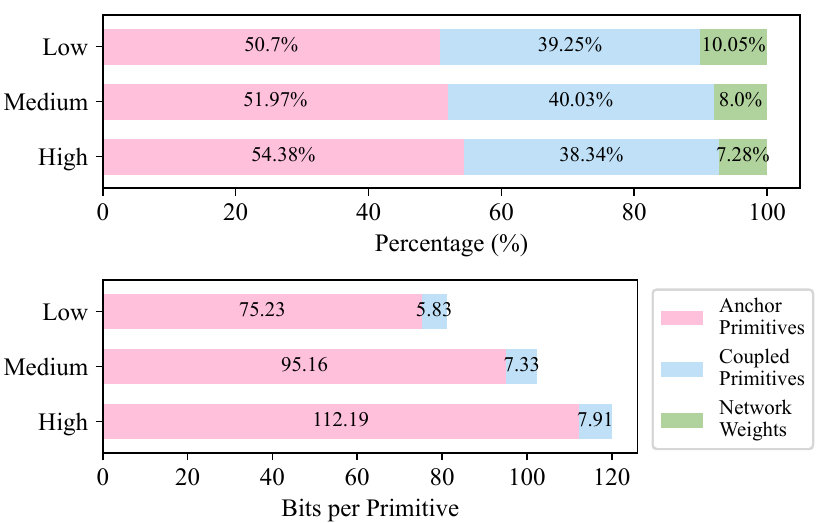}
  \caption{Bitstream analysis on the \textit{Train} scene. \textbf{Upper}: Proportion of components within the bitstream. \textbf{Bottom}: Average bits consumption of anchor primitives and coupled primitives.}
  \label{fig::rate_analysis_static}
\end{figure}

\textbf{Effectiveness on spatial primitive prediction.}
Our spatial primitive prediction module is devised to enhance the compactness of scene representations by effectively removing spatial redundancy of 3D Gaussians.
This is adeptly achieved by predicting 3D Gaussians using reference information from a minimal set of anchor primitives and residual information from coupled primitives. 
To validate this predictive paradigm, we incorporate it into the baseline 3DGS~\cite{kerbl20233d} and conduct ablation studies using the Tanks\&Temples dataset~\cite{knapitsch2017tanks}.
As shown in Table~\ref{tab::ablation_studies}, our spatial primitive prediction module significantly improves the compactness of scene representations, achieving size reductions of 177.56~MB and 563.97~MB for the \textit{Train} and \textit{Truck} scenes, respectively.
This is because the spatial primitive prediction module can eliminate spatial redundancies by exploiting inter-primitive correlations, thereby improving the compactness of representation.
Furthermore, to intuitively demonstrate the efficacy of the spatial primitive prediction, we analyze the bitstream composition for the \textit{Train} scene.
As illustrated in Fig.~\ref{fig::rate_analysis_static}, despite the coupled primitives significantly outnumbering the anchor primitives, their total rate consumption remains comparable. 
Particularly, the per-primitive bitrate of coupled primitives is approximately one-tenth that of anchor primitives.
This is because the proposed prediction paradigm enables coupled primitives to be succinctly represented in residual form and facilitates the compactness of scene representations.

\begin{table*}
\caption{Ablation studies for components within the proposed spatial primitive prediction}
\label{tab::ab_spatial_pred}
\centering
\begin{tabular}{cc|cccc|cccc}
\toprule
\multicolumn{1}{c}{Residual} & \multicolumn{1}{c|}{Spatial}  & \multicolumn{4}{c|}{\textit{Train}}  & \multicolumn{4}{c}{\textit{Truck}}  \\
\cmidrule{3-10} \multicolumn{1}{c}{Embeddings}  & \multicolumn{1}{c|}{Context}  & \multicolumn{1}{c}{PSNR (dB)} & \multicolumn{1}{c}{SSIM} & \multicolumn{1}{c}{LPIPS} & \multicolumn{1}{c|}{Size (MB)} & \multicolumn{1}{c}{PSNR (dB)} & \multicolumn{1}{c}{SSIM} & \multicolumn{1}{c}{LPIPS} & \multicolumn{1}{c}{Size (MB)}  \\
\midrule
$\times$      & $\checkmark$ & 21.87 & 0.80 & 0.24 & 4.53 & 24.70  & 0.85 & 0.20 & 3.16 \\
$\checkmark$  & $\times$     & 22.29 & 0.81 & 0.23 & 5.45 & 25.07 & 0.86 & 0.19 & 3.43 \\
$\checkmark$  & $\checkmark$ & 22.44 & 0.81 & 0.23 & 3.91 & 25.09 & 0.86 & 0.20 & 3.09 \\
\bottomrule
\end{tabular}
\end{table*}

Subsequently, we conduct ablation studies to examine the effectiveness of components within the spatial primitive prediction module. 
First, we assess the efficacy of residual embeddings within coupled primitives.
We craft a variant based on the Scaffold-GS~\cite{lu20243d} primitive derivation paradigm, which relies solely on anchor primitive embeddings for prediction.
As shown in Table~\ref{tab::ab_spatial_pred}, this variant experiences significant quality degradation, with decreases of 0.57~dB and 0.39~dB for the \textit{Train} and \textit{Truck} scenes, respectively.
These results underscore the importance of residual embeddings in capturing unique characteristics of individual 3D Gaussians.
Meanwhile, we validate the effectiveness of the spatial context in our prediction module. 
Results shown in Table~\ref{tab::ab_spatial_pred} demonstrate that removing spatial context leads to performance degradation, highlighting the importance of exploring spatial correlations between anchor primitives during the prediction.
Furthermore, we investigate the optimal ratio $K$ of coupled primitives to anchor primitives. 
As shown in Table~\ref{tab::diff_k}, the model achieves optimal performance with $K=10$, which we subsequently adopted for our experiments. 
Notably, increasing $K$ to 15 results in performance degradation, likely because excessive combination of coupled primitive could diminish inter-primitive correlations and adversely affect prediction accuracy.

\textbf{Effectiveness on rate-constrained optimization.}
The rate-constrained optimization module is crafted to improve the compactness of primitive parameters by jointly optimizing rendering quality and coding costs. 
To assess this module, we incorporate it with the spatial primitive prediction module, and establish our complete method for compact representation of static scenes.
The results in Table~\ref{tab::ablation_studies} demonstrate the significant impact of our rate-constrained optimization module. 
For the \textit{Train} scene, implementing rate-constrained optimization reduces the model size from 79.88~MB to 3.91~MB.
Similarly, results on the \textit{Truck} scene exhibit a remarkable 93.47\% reduction in size.
These substantial improvements in compression efficiency stem from the rate-constrained optimization that enables the proposed method to learn compact primitive representations.

\begin{table*}[t]
\caption{Ablation studies on the proportion of coupled primitives}
\label{tab::diff_k}
\centering
\begin{tabular}{l|cccc|cccc}
\toprule
& \multicolumn{4}{c|}{\textit{Train}} & \multicolumn{4}{c}{\textit{Truck}} \\
\cmidrule{2-9} \multicolumn{1}{c|}{} & PSNR (dB) & SSIM & LPIPS & Size (MB) & PSNR (dB) & SSIM & LPIPS & Size (MB)\\
\midrule
$K=5$  & 22.46 & 0.81 & 0.23 & 4.52 & 24.89 & 0.86 & 0.20 & 3.42 \\
$K=10$ & 22.44 & 0.81 & 0.23 & 3.91 & 25.09 & 0.86 & 0.20 & 3.09 \\
$K=15$ & 22.26 & 0.81 & 0.23 & 4.87 & 24.79 & 0.85 & 0.20 & 3.09 \\
\bottomrule
\end{tabular}
\end{table*}

\begin{table*}
\caption{Ablation studies for components within the proposed rate-constrained optimization}
\label{tab::ab_rate_opt}
\centering
\begin{tabular}{cc|cccc|cccc}
\toprule
\multicolumn{1}{c}{Hyper} & \multicolumn{1}{c|}{Spatial}  & \multicolumn{4}{c|}{\textit{Train}}  & \multicolumn{4}{c}{\textit{Truck}}  \\
\cmidrule{3-10} \multicolumn{1}{c}{Priors}  & \multicolumn{1}{c|}{Priors}  & \multicolumn{1}{c}{PSNR (dB)} & \multicolumn{1}{c}{SSIM} & \multicolumn{1}{c}{LPIPS} & \multicolumn{1}{c|}{Size (MB)} & \multicolumn{1}{c}{PSNR (dB)} & \multicolumn{1}{c}{SSIM} & \multicolumn{1}{c}{LPIPS} & \multicolumn{1}{c}{Size (MB)}  \\
\midrule
$\times$     & $\checkmark$ & 22.36 & 0.81 & 0.23 & 4.30 & 25.04 & 0.86 & 0.20 & 3.38 \\
$\checkmark$ & $\times$     & 22.41 & 0.81 & 0.23 & 4.15 & 25.11 & 0.86 & 0.19 & 3.76 \\
$\checkmark$ & $\checkmark$ & 22.44 & 0.81 & 0.23 & 3.91 & 25.09 & 0.86 & 0.20 & 3.09  \\
\bottomrule
\end{tabular}
\end{table*}

Moreover, we conduct ablations on the entropy model within the rate-constrained optimization module. 
Specifically, we examine the efficacy of hyperpriors by creating a variant that relies solely on spatial context for probability modeling.
As shown in Table~\ref{tab::ab_rate_opt}, removing hyperpriors increases the model size from 3.91~MB to 4.30~MB for the \textit{Train} scene and from 3.09~MB to 3.38~MB for the \textit{Truck} scene. 
This performance drop occurs because this variant loses access to critical prior information pertinent to the parameters being encoded, demonstrating that spatial priors alone are insufficient for probability modeling.
We also investigate the importance of spatial priors within the entropy model by developing a variant without such priors.
It can be observed from Table~\ref{tab::ab_rate_opt} that the removal of spatial priors negatively impacts compression performance, invoking size increases of 6.14\% and 21.68\% for the \textit{Train} and \textit{Truck} scenes, respectively.
These findings underscore the significance of both hyperpriors and spatial priors in achieving optimal compression efficiency.

\textbf{Effectiveness on temporal primitive prediction.}
Our temporal primitive prediction module is proposed to utilize primitives from preceding scenes as references to predict those required for current scene construction, enabling compact representation of dynamic scenes. 
To validate its effectiveness, we conduct ablation studies using the \textit{Coffee Martini} and \textit{Sear Steak} scenes from the Neural 3D Video dataset~\cite{li2022neural}, comparing against the baseline that maintains inherited primitives without modification.
As shown in Table~\ref{tab::ablation_dynamic}, our temporal prediction module significantly improves modeling efficiency, achieving rendering quality improvements of 3.51~dB and 5.09~dB on the \textit{Coffee Martini} and \textit{Sear Steak} scenes, respectively, while maintaining comparable model sizes.
These improvements stem from the ability of this module to efficiently adapt inherited primitives to current scenes by using compact temporal residues with minimal rate consumption.

To further validate the temporal primitive prediction, we conduct detailed component ablation studies. 
First, we investigate the efficacy of our static-dynamic primitive disentanglement strategy by comparing it against a variant that treats all inherited primitives as dynamic ones. 
As demonstrated in Table~\ref{tab::ab_temporal_pred}, this variant experiences substantial quality degradation, with decreases of 9.59~dB and 14.57~dB for the \textit{Coffee Martini} and \textit{Sear Steak} scenes, respectively.
This degradation occurs because forcing static region primitives to learn non-existent motion biases not only compromises reconstruction quality in static regions but also affects modeling accuracy in dynamic regions.
Additionally, we examine the effectiveness of global illumination compensation for static primitives. 
Results in Table~\ref{tab::ab_temporal_pred} show that removing such compensation leads to quality reductions of 0.56~dB and 0.53~dB for the \textit{Coffee Martini} and \textit{Sear Steak} scenes, respectively, underscoring the importance of this component in maintaining rendering quality.

\begin{table*}
\caption{Ablation studies on the Neural 3D Video dataset~\cite{li2022neural}}
\label{tab::ablation_dynamic}
\centering
\resizebox{\linewidth}{!}{
\begin{tabular}{cc|ccc|ccc}
\toprule
\multicolumn{1}{c}{Temporal Primitive} & \multicolumn{1}{c|}{Temporal}  & \multicolumn{3}{c|}{\textit{Coffee Martini}}  & \multicolumn{3}{c}{\textit{Sear Steak}}  \\
\cmidrule{3-8} \multicolumn{1}{c}{Prediction}  & \multicolumn{1}{c|}{Adaptive Control}  & \multicolumn{1}{c}{PSNR (dB)} & \multicolumn{1}{c}{Total Size (MB)} & \multicolumn{1}{c|}{Per Frame Size (MB)} & \multicolumn{1}{c}{PSNR (dB)} & \multicolumn{1}{c}{Total Size (MB)} & \multicolumn{1}{c}{Per Frame Size (MB)} \\
\midrule
$\times$      & $\times$      & 24.86 & 14.20 & 0.05 & 27.86 & 20.43 & 0.07 \\
$\checkmark$  & $\times$      & 28.67 & 20.45 & 0.07 & 32.90 & 28.07 & 0.09 \\
$\checkmark$  & $\checkmark$  & 28.37 & 13.79 & 0.05 & 32.95 & 16.39 & 0.05 \\
\bottomrule
\end{tabular}}
\end{table*}

\begin{table*}
\caption{Ablation studies for components within the proposed temporal primitive prediction}
\label{tab::ab_temporal_pred}
\centering
\resizebox{\linewidth}{!}{
\begin{tabular}{cc|ccc|ccc}
\toprule
\multicolumn{1}{c}{Primitive} & \multicolumn{1}{c|}{Appearance}  & \multicolumn{3}{c|}{\textit{Coffee Martini}}  & \multicolumn{3}{c}{\textit{Sear Steak}}  \\
\cmidrule{3-8} \multicolumn{1}{c}{Disentanglement}  & \multicolumn{1}{c|}{Compensation}  & \multicolumn{1}{c}{PSNR (dB)} & \multicolumn{1}{c}{Total Size (MB)} & \multicolumn{1}{c|}{Per Frame Size (MB)} & \multicolumn{1}{c}{PSNR (dB)} & \multicolumn{1}{c}{Total Size (MB)} & \multicolumn{1}{c}{Per Frame Size (MB)} \\
\midrule
$\times$      & $\checkmark$  & 18.78 & 49.44 & 0.16 & 18.38 & 31.79 & 0.11 \\
$\checkmark$  & $\times$      & 27.81 & 17.05 & 0.06 & 32.42 & 18.54 & 0.06 \\
$\checkmark$  & $\checkmark$  & 28.37 & 13.79 & 0.05 & 32.95 & 16.39 & 0.05 \\
\bottomrule
\end{tabular}}
\end{table*}

\textbf{Effectiveness on temporal adaptive control.}
The temporal adaptive control module is proposed to adaptively manage the transition between dynamic and static primitives during optimization while introducing new primitives as needed. 
To evaluate its effectiveness, we integrate it with the temporal primitive prediction module and establish our complete method for compact dynamic scene representation.
The corresponding experimental results are depicted in Table~\ref{tab::ablation_dynamic}. 
It can be observed that our temporal adaptive control module significantly reduces the model size from 20.45~MB to 13.79~MB for the \textit{Coffee Martini} scene, and from 28.07~MB to 16.39~MB for the \textit{Sear Steak} scene.
This is because this module can mitigate initial disentanglement misclassification and improve the modeling accuracy for both static and dynamic primitives.
This advancement further improves the overall performance of rendering quality and coding costs.

\subsection{Complexity Analysis}
Table~\ref{tab::complexity} reports the complexity comparison between our proposed method and existing compression methods~\cite{navaneet2025compgs, niedermayr2024compressed, lee2024compact, girish2024eagles, liu2024compgs, chen2024hac} on the Tanks\&Temples dataset~\cite{knapitsch2017tanks}.
Our method requires a training time of 28.40 minutes, falling between faster methods~\cite{navaneet2025compgs, niedermayr2024compressed, girish2024eagles, chen2024hac} and slower methods~\cite{lee2024compact, liu2024compgs}. 
In terms of compression complexity, our method achieves efficient encoding and decoding, with both consuming less than 10 seconds.
Owing to the highly parallel splatting rendering algorithm~\cite{zwicker2002ewa}, the average rendering time of our method averages 5.86 milliseconds, which is comparable to existing methods.
Moreover, the proposed method exhibits improved computational efficiency compared to our previous work~\cite{liu2024compgs}, with reduced training, encoding, and decoding times.
This enhancement stems from our streamlined entropy model design, which eliminates unnecessary conditional dependencies and obtains a lower computational overhead while maintaining compression effectiveness.

\section{Conclusion}
This paper proposes Compressed Gaussian Splatting (CompGS++), a novel scheme that leverages compact Gaussian primitive representations to achieve efficient 3D scene modeling and remarkable size reduction.
Specifically, we devise a spatial primitive prediction module to eliminate spatial primitive redundancy. 
By establishing predictive relationships between primitives, our framework enables efficient representation of most 3D Gaussians as residuals, substantially reducing spatial redundancy in scene representations.
Moreover, we elevate our prediction paradigm to dynamic scenes through the temporal primitive prediction module. 
This module leverages primitive correlations across timestamps and compactly constructs the current scenes by referencing previous primitives, thereby effectively reducing temporal redundancy.
To further improve compression efficiency, we propose the rate-constrained optimization module that addresses parameter redundancy within each primitive. 
This module optimizes for the trade-off between reconstruction quality and rate consumption, resulting in more compact primitive parameters while maintaining high rendering quality.
Through the integration of these sophisticated modules, our CompGS++ demonstrates superior performance compared to existing compression methods, achieving exceptional compression ratios while preserving high-quality rendering capabilities.

\begin{table}[t]
\caption{Complexity comparison on the Tanks\&Temples dataset~\cite{knapitsch2017tanks}}
\centering
\resizebox{0.9\linewidth}{!}{
\begin{tabular}{l|rrrr}
\toprule
 & \multicolumn{1}{c}{Train} & \multicolumn{1}{c}{Enc-time} & \multicolumn{1}{c}{Dec-time} & \multicolumn{1}{c}{Render} \\
 & \multicolumn{1}{c}{(min)} & \multicolumn{1}{c}{(s)} & \multicolumn{1}{c}{(s)} & \multicolumn{1}{c}{(ms)} \\
\midrule
Navaneet et~al.~\cite{navaneet2025compgs}          & 14.38 & 68.29 & 12.32 & 9.88 \\
Niedermayr et~al.~\cite{niedermayr2024compressed}  & 15.50 &  2.23 &  0.25 & 9.74 \\
Lee et~al.~\cite{lee2024compact}                   & 44.70 &  1.96 &  0.18 & 6.60 \\
Girish et~al.~\cite{girish2024eagles}              &  8.95 &  0.54 &  0.64 & 6.96 \\
Liu et~al.~\cite{liu2024compgs}                    & 37.83 &  6.27 &  4.46 & 5.32 \\
Chen et~al.~\cite{chen2024hac}                     & 20.05 & 35.36 & 31.97 & 6.95 \\
Proposed                                           & 28.40 &  4.35 &  3.87 & 5.86 \\
\bottomrule
\end{tabular}}
\label{tab::complexity}
\end{table}

{\small
\bibliographystyle{IEEEtran}
\bibliography{reference}

\begin{thebibliography}{10}
\providecommand{\url}[1]{#1}
\csname url@samestyle\endcsname
\providecommand{\newblock}{\relax}
\providecommand{\bibinfo}[2]{#2}
\providecommand{\BIBentrySTDinterwordspacing}{\spaceskip=0pt\relax}
\providecommand{\BIBentryALTinterwordstretchfactor}{4}
\providecommand{\BIBentryALTinterwordspacing}{\spaceskip=\fontdimen2\font plus
\BIBentryALTinterwordstretchfactor\fontdimen3\font minus \fontdimen4\font\relax}
\providecommand{\BIBforeignlanguage}[2]{{%
\expandafter\ifx\csname l@#1\endcsname\relax
\typeout{** WARNING: IEEEtran.bst: No hyphenation pattern has been}%
\typeout{** loaded for the language `#1'. Using the pattern for}%
\typeout{** the default language instead.}%
\else
\language=\csname l@#1\endcsname
\fi
#2}}
\providecommand{\BIBdecl}{\relax}
\BIBdecl

\bibitem{kerbl20233d}
B.~Kerbl, G.~Kopanas, T.~Leimk{\"u}hler, and G.~Drettakis, ``{3D} {Gaussian} splatting for real-time radiance field rendering,'' \emph{ACM Transactions on Graphics}, vol.~42, no.~4, pp. 1--14, 2023.

\bibitem{lei2025gaussnav}
X.~Lei, M.~Wang, W.~Zhou, and H.~Li, ``{GaussNav}: {Gaussian} splatting for visual navigation,'' \emph{IEEE Transactions on Pattern Analysis and Machine Intelligence}, pp. 1--14, 2025.

\bibitem{qu2024z}
Z.~Qu, O.~Vengurlekar, M.~Qadri, K.~Zhang, M.~Kaess, C.~Metzler, S.~Jayasuriya, and A.~Pediredla, ``{Z-Splat}: Z-axis {Gaussian} splatting for camera-sonar fusion,'' \emph{IEEE Transactions on Pattern Analysis and Machine Intelligence}, pp. 1--12, 2024.

\bibitem{yin2025ms}
Z.-X. Yin, P.-Y. Jiao, J.~Qiu, M.-M. Cheng, and B.~Ren, ``{MS-NeRF}: Multi-space neural radiance fields,'' \emph{IEEE Transactions on Pattern Analysis and Machine Intelligence}, pp. 1--18, 2025.

\bibitem{gableman2024incorporating}
M.~Gableman and A.~Kak, ``Incorporating season and solar specificity into renderings made by a {NeRF} architecture using satellite images,'' \emph{IEEE Transactions on Pattern Analysis and Machine Intelligence}, vol.~46, no.~6, pp. 4348--4365, 2024.

\bibitem{chen2024s}
Y.~Chen, J.~Zhang, Z.~Xie, W.~Li, F.~Zhang, J.~Lu, and L.~Zhang, ``{S-NeRF++}: Autonomous driving simulation via neural reconstruction and generation,'' \emph{IEEE Transactions on Pattern Analysis and Machine Intelligence}, pp. 1--19, 2025.

\bibitem{ramirez2024deep}
P.~Z. Ramirez, L.~De~Luigi, D.~Sirocchi, A.~Cardace, R.~Spezialetti, F.~Ballerini, S.~Salti, and L.~Di~Stefano, ``Deep learning on object-centric {3D} neural fields,'' \emph{IEEE Transactions on Pattern Analysis and Machine Intelligence}, vol.~46, no.~12, pp. 9940--9956, 2024.

\bibitem{mildenhall2021nerf}
B.~Mildenhall, P.~P. Srinivasan, M.~Tancik, J.~T. Barron, R.~Ramamoorthi, and R.~Ng, ``{NeRF}: Representing scenes as neural radiance fields for view synthesis,'' \emph{Communications of the ACM}, vol.~65, no.~1, pp. 99--106, 2021.

\bibitem{zwicker2002ewa}
M.~Zwicker, H.~Pfister, J.~Van~Baar, and M.~Gross, ``{EWA} splatting,'' \emph{IEEE Transactions on Visualization and Computer Graphics}, vol.~8, no.~3, pp. 223--238, 2002.

\bibitem{huang20242d}
B.~Huang, Z.~Yu, A.~Chen, A.~Geiger, and S.~Gao, ``{2D} {Gaussian} splatting for geometrically accurate radiance fields,'' in \emph{Proceedings of the ACM SIGGRAPH Conference}, 2024, pp. 1--11.

\bibitem{hamdi2024ges}
A.~Hamdi, L.~Melas-Kyriazi, J.~Mai, G.~Qian, R.~Liu, C.~Vondrick, B.~Ghanem, and A.~Vedaldi, ``{GES}: Generalized exponential splatting for efficient radiance field rendering,'' in \emph{Proceedings of the IEEE/CVF Conference on Computer Vision and Pattern Recognition}, 2024, pp. 19\,812--19\,822.

\bibitem{wang2024tangram}
Y.~Wang, N.~Zhong, M.~Chen, L.~Wang, and Y.~Guo, ``Tangram-splatting: Optimizing {3D} {Gaussian} splatting through tangram-inspired shape priors,'' in \emph{Proceedings of the ACM Multimedia}, 2024, pp. 3075--3083.

\bibitem{li20243d}
H.~Li, J.~Liu, M.~Sznaier, and O.~Camps, ``{3D-HGS}: {3D} half-{Gaussian} splatting,'' \emph{arXiv:2406.02720}, pp. 1--11, 2024.

\bibitem{qu2024disc}
H.~Qu, Z.~Li, H.~Rahmani, Y.~Cai, and J.~Liu, ``{DisC-GS}: Discontinuity-aware {Gaussian} splatting,'' in \emph{Proceedings of the Advances in Neural Information Processing Systems}, vol.~37, 2024, pp. 112\,284--112\,309.

\bibitem{wu20244d}
G.~Wu, T.~Yi, J.~Fang, L.~Xie, X.~Zhang, W.~Wei, W.~Liu, Q.~Tian, and X.~Wang, ``{4D} {Gaussian} splatting for real-time dynamic scene rendering,'' in \emph{Proceedings of the IEEE/CVF Conference on Computer Vision and Pattern Recognition}, 2024, pp. 20\,310--20\,320.

\bibitem{kratimenos2023dynmf}
A.~Kratimenos, J.~Lei, and K.~Daniilidis, ``{DynMF}: Neural motion factorization for real-time dynamic view synthesis with {3D} {Gaussian} splatting,'' in \emph{Proceedings of the European Conference on Computer Vision}, 2024, pp. 252--269.

\bibitem{li2024spacetime}
Z.~Li, Z.~Chen, Z.~Li, and Y.~Xu, ``Spacetime {Gaussian} feature splatting for real-time dynamic view synthesis,'' in \emph{Proceedings of the IEEE/CVF Conference on Computer Vision and Pattern Recognition}, 2024, pp. 8508--8520.

\bibitem{yang2023real}
Z.~Yang, H.~Yang, Z.~Pan, and L.~Zhang, ``Real-time photorealistic dynamic scene representation and rendering with {4D} {Gaussian} splatting,'' in \emph{Proceedings of the International Conference on Learning Representations}, 2024, pp. 1--12.

\bibitem{luiten2024dynamic}
J.~Luiten, G.~Kopanas, B.~Leibe, and D.~Ramanan, ``Dynamic {3D} {Gaussians}: Tracking by persistent dynamic view synthesis,'' in \emph{Proceedings of the International Conference on 3D Vision}, 2024, pp. 800--809.

\bibitem{sun20243dgstream}
J.~Sun, H.~Jiao, G.~Li, Z.~Zhang, L.~Zhao, and W.~Xing, ``{3DGStream}: On-the-fly training of {3D} {Gaussians} for efficient streaming of photo-realistic free-viewpoint videos,'' in \emph{Proceedings of the IEEE/CVF Conference on Computer Vision and Pattern Recognition}, 2024, pp. 20\,675--20\,685.

\bibitem{gao2024hicom}
Q.~Gao, J.~Meng, C.~Wen, J.~Chen, and J.~Zhang, ``{HiCoM}: Hierarchical coherent motion for dynamic streamable scenes with {3D} {Gaussian} splatting,'' in \emph{Proceedings of the Advances in Neural Information Processing Systems}, vol.~37, 2024, pp. 80\,609--80\,633.

\bibitem{navaneet2025compgs}
K.~Navaneet, K.~Pourahmadi~Meibodi, S.~Abbasi~Koohpayegani, and H.~Pirsiavash, ``{CompGS}: Smaller and faster {Gaussian} splatting with vector quantization,'' in \emph{Proceedings of the European Conference on Computer Vision}, 2025, pp. 330--349.

\bibitem{niedermayr2024compressed}
S.~Niedermayr, J.~Stumpfegger, and R.~Westermann, ``Compressed {3D} {Gaussian} splatting for accelerated novel view synthesis,'' in \emph{Proceedings of the IEEE/CVF Conference on Computer Vision and Pattern Recognition}, 2024, pp. 10\,349--10\,358.

\bibitem{lee2024compact}
J.~C. Lee, D.~Rho, X.~Sun, J.~H. Ko, and E.~Park, ``Compact {3D} {Gaussian} representation for radiance field,'' in \emph{Proceedings of the IEEE/CVF Conference on Computer Vision and Pattern Recognition}, 2024, pp. 21\,719--21\,728.

\bibitem{girish2024eagles}
S.~Girish, K.~Gupta, and A.~Shrivastava, ``{EAGLES}: Efficient accelerated {3D} gaussians with lightweight encodings,'' in \emph{Proceedings of the European Conference on Computer Vision}, 2024, pp. 54--71.

\bibitem{liu2024compgs}
X.~Liu, X.~Wu, P.~Zhang, S.~Wang, Z.~Li, and S.~Kwong, ``{CompGS}: Efficient {3D} scene representation via compressed {Gaussian} splatting,'' in \emph{Proceedings of the ACM International Conference on Multimedia}, 2024, pp. 2936--2944.

\bibitem{chen2024hac}
Y.~Chen, Q.~Wu, W.~Lin, M.~Harandi, and J.~Cai, ``{HAC}: Hash-grid assisted context for {3D} {Gaussian} splatting compression,'' in \emph{Proceedings of the European Conference on Computer Vision}, 2024, pp. 422--438.

\bibitem{ye2024absgs}
Z.~Ye, W.~Li, S.~Liu, P.~Qiao, and Y.~Dou, ``{AbsGS}: Recovering fine details in {3D} {Gaussian} splatting,'' in \emph{Proceedings of the ACM Multimedia}, 2024, pp. 1053--1061.

\bibitem{bulo2024revising}
S.~R. Bul{\`o}, L.~Porzi, and P.~Kontschieder, ``Revising densification in {Gaussian} splatting,'' in \emph{Proceedings of the European Conference on Computer Vision}, 2024, pp. 347--362.

\bibitem{zhang2024pixel}
Z.~Zhang, W.~Hu, Y.~Lao, T.~He, and H.~Zhao, ``Pixel-{GS}: Density control with pixel-aware gradient for {3D} {Gaussian} splatting,'' in \emph{Proceedings of the European Conference on Computer Vision}, 2024, pp. 326--342.

\bibitem{wang2024adr}
X.~Wang, R.~Yi, and L.~Ma, ``{AdR}-{Gaussian}: Accelerating {Gaussian} splatting with adaptive radius,'' in \emph{Proceedings of the ACM SIGGRAPH Asia Conference}, 2024, pp. 1--10.

\bibitem{huang2024gs++}
L.~Huang, J.~Bai, J.~Guo, and Y.~Guo, ``On the error analysis of {3D} {Gaussian} splatting and an optimal projection strategy,'' in \emph{Proceedings of the European Conference on Computer Vision}, 2024, pp. 247--263.

\bibitem{fan2024lightgaussian}
Z.~Fan, K.~Wang, K.~Wen, Z.~Zhu, D.~Xu, and Z.~Wang, ``{LightGaussian}: Unbounded {3D} {Gaussian} compression with 15x reduction and 200+ {FPS},'' in \emph{Proceedings of the Advances in Neural Information Processing Systems}, 2023, pp. 1--14.

\bibitem{balle2018variational}
J.~Ball{\'e}, D.~Minnen, S.~Singh, S.~J. Hwang, and N.~Johnston, ``Variational image compression with a scale hyperprior,'' in \emph{Proceedings of the International Conference on Learning Representations}, 2018, pp. 1--13.

\bibitem{xu20244k4d}
Z.~Xu, S.~Peng, H.~Lin, G.~He, J.~Sun, Y.~Shen, H.~Bao, and X.~Zhou, ``{4K4D}: Real-time {4D} view synthesis at {4K} resolution,'' in \emph{Proceedings of the IEEE/CVF Conference on Computer Vision and Pattern Recognition}, 2024, pp. 20\,029--20\,040.

\bibitem{knapitsch2017tanks}
A.~Knapitsch, J.~Park, Q.-Y. Zhou, and V.~Koltun, ``Tanks and temples: Benchmarking large-scale scene reconstruction,'' \emph{ACM Transactions on Graphics}, vol.~36, no.~4, pp. 1--13, 2017.

\bibitem{hedman2018deep}
P.~Hedman, J.~Philip, T.~Price, J.-M. Frahm, G.~Drettakis, and G.~Brostow, ``Deep blending for free-viewpoint image-based rendering,'' \emph{ACM Transactions on Graphics}, vol.~37, no.~6, pp. 1--15, 2018.

\bibitem{schwarz2018emerging}
S.~Schwarz, M.~Preda, V.~Baroncini, M.~Budagavi, P.~Cesar, P.~A. Chou, R.~A. Cohen, M.~Krivoku{\'c}a, S.~Lasserre, Z.~Li \emph{et~al.}, ``Emerging {MPEG} standards for point cloud compression,'' \emph{IEEE Journal on Emerging and Selected Topics in Circuits and Systems}, vol.~9, no.~1, pp. 133--148, 2018.

\bibitem{mentzer2019practical}
F.~Mentzer, E.~Agustsson, M.~Tschannen, R.~Timofte, and L.~V. Gool, ``Practical full resolution learned lossless image compression,'' in \emph{Proceedings of the IEEE/CVF Conference on Computer Vision and Pattern Recognition}, 2019, pp. 10\,629--10\,638.

\bibitem{barron2022mip}
J.~T. Barron, B.~Mildenhall, D.~Verbin, P.~P. Srinivasan, and P.~Hedman, ``Mip-{NeRF} 360: Unbounded anti-aliased neural radiance fields,'' in \emph{Proceedings of the IEEE/CVF Conference on Computer Vision and Pattern Recognition}, 2022, pp. 5470--5479.

\bibitem{wang2004image}
Z.~Wang, A.~C. Bovik, H.~R. Sheikh, and E.~P. Simoncelli, ``Image quality assessment: From error visibility to structural similarity,'' \emph{IEEE Transactions on Image Processing}, vol.~13, no.~4, pp. 600--612, 2004.

\bibitem{zhang2018unreasonable}
R.~Zhang, P.~Isola, A.~A. Efros, E.~Shechtman, and O.~Wang, ``The unreasonable effectiveness of deep features as a perceptual metric,'' in \emph{Proceedings of the IEEE Conference on Computer Vision and Pattern Recognition}, 2018, pp. 586--595.

\bibitem{li2022neural}
T.~Li, M.~Slavcheva, M.~Zollhoefer, S.~Green, C.~Lassner, C.~Kim, T.~Schmidt, S.~Lovegrove, M.~Goesele, R.~Newcombe \emph{et~al.}, ``Neural {3D} video synthesis from multi-view video,'' in \emph{Proceedings of the IEEE/CVF Conference on Computer Vision and Pattern Recognition}, 2022, pp. 5521--5531.

\bibitem{lu20243d}
Z.~Lu, X.~Guo, L.~Hui, T.~Chen, M.~Yang, X.~Tang, F.~Zhu, and Y.~Dai, ``{3D} geometry-aware deformable {Gaussian} splatting for dynamic view synthesis,'' in \emph{Proceedings of the IEEE/CVF Conference on Computer Vision and Pattern Recognition}, 2024, pp. 8900--8910.

\end{thebibliography}
}

\end{document}